\documentclass[conference]{IEEEtran}
\IEEEoverridecommandlockouts
\usepackage{comment}
\usepackage{float}
\usepackage{cite}
\usepackage{hyperref}
\usepackage{amsmath,amssymb,amsfonts}
\usepackage{algorithmic}
\usepackage{graphicx}
\usepackage{textcomp}
\usepackage{xcolor}
\usepackage{subcaption}
\usepackage{multirow}
\usepackage{adjustbox}
\usepackage{booktabs}
\usepackage{color,soul}
\usepackage[ruled,linesnumbered]{algorithm2e}
\usepackage{tabularx}
    \newcolumntype{L}{>{\raggedright\arraybackslash}X}
    
\def\BibTeX{{\rm B\kern-.05em{\sc i\kern-.025em b}\kern-.08em
    T\kern-.1667em\lower.7ex\hbox{E}\kern-.125emX}}

\SetKwRepeat{Do}{do}{while}

\begin{document}

\title{Mask Off: Analytic-based Malware Detection By Transfer Learning and Model Personalization 
\\
}

\author{\IEEEauthorblockN{Amirmohammad Pasdar}
\IEEEauthorblockA{\textit{School of Computing} \\
\textit{Macquarie University}\\
Sydney, Australia \\
amirmohammad.pasdar@hdr.mq.edu.au}
\and
\IEEEauthorblockN{Young Choon Lee}
\IEEEauthorblockA{\textit{School of Computing} \\
\textit{Macquarie University}\\
Sydney, Australia \\
young.lee@mq.edu.au}
\and
\IEEEauthorblockN{Seok-Hee Hong}
\IEEEauthorblockA{\textit{School of Computer Science} \\
\textit{University of Sydney}\\
Sydney, Australia \\
seokhee.hong@sydney.edu.au}
}

\maketitle

\begin{abstract}
The vulnerability of smartphones to cyberattacks has been a severe concern to users arising from the integrity of installed applications (\textit{apps}). Although applications are to provide legitimate and diversified on-the-go services, harmful and dangerous ones have also uncovered the feasible way to penetrate smartphones for malicious behaviors. Thorough application analysis is key to revealing malicious intent and providing more insights into the application behavior for security risk assessments. Such in-depth analysis motivates employing deep neural networks (DNNs) for a set of features and patterns extracted from applications to facilitate detecting potentially dangerous applications independently. This paper presents an Analytic-based deep neural network, Android Malware detection (ADAM), that employs a fine-grained set of features to train feature-specific DNNs to have consensus on the application labels when their ground truth is unknown. In addition, ADAM leverages the transfer learning technique to obtain its adjustability to new applications across smartphones for recycling the pre-trained model(s) and making them more adaptable by model personalization and federated learning techniques. This adjustability is also assisted by federated learning guards, which protect ADAM against poisoning attacks through model analysis. ADAM relies on a diverse dataset containing more than 153000 applications with over 41000 extracted features for DNNs training. The ADAM's feature-specific DNNs, on average, achieved more than 98\% accuracy, resulting in an outstanding performance against data manipulation attacks.

\end{abstract}


\section{Introduction}

Smartphones have indeed changed people's daily life. According to Statista \cite{statista}, the number of smartphone users exceeds six billion, and the world will witness substantial growth in the next few years. This growth has resulted in an explosion of numerous application development for various personal or business on-the-go services on smartphones. Although promising, harmful, and potentially dangerous applications (i.e., malware) have also discovered the feasible way to infiltrate smartphones for malicious behaviors. The Nokia Threat Intelligence Report \cite{nokiareport} indicated the infection rate of Internet of Things (IoT) devices doubled in 2020 compared to the metrics reported in 2019, reaching an alarming level of 33\% of the infectious rate. 

Due to its openness, the Android operating system (OS) has approximately 80\% of the worldwide market share and is the most frequently used OS on different mobile platforms. Therefore, many developers are attracted to Android, and many applications are exponentially becoming available for smartphones. These applications are shared through the official market store (i.e., Google Play Store) and non-official repositories worldwide, which help users to discover and install the applications easily. Even though the official market store applies essential application development and distribution regulations (i.e., Play Protect \cite{playprotect}), malware is still spread in the market (e.g., \cite{jokermalware1}). The non-official ones have also become the potential gateway for malware distribution. Hence, it is still vital to thoroughly study Android applications to have their behaviors profiled for revealing dangerous and harmful applications.

Android malware detection has been exhaustively studied (e.g., \cite{power1,power3,ml2,cloud2,cloud4,cloud5,network6,api_access}) with a primary focus on a small set of information collected in virtualized environments \cite{maldroid} or through different tools (e.g., \cite{drozer,andromaly}). While dynamic analysis has also been studied, it necessitates to \emph{unroot} the Android OS utilizing underlying OS features, which is another potential risk to smartphones. Considering these features and the maturity of Android OS \cite{privacyandroid}, malware can still evade detection using classifiers and machine learning (ML) techniques. Such an approach primarily does not take an in-depth analysis of applications in conjunction with factors such as profiling power consumption or abnormal communications (e.g., \cite{ml5,ml7,ml10,ml12,ml13,ml14, analysis4}). Moreover, real-time detection by anti-malware or antivirus apps is not feasible as they rely on central signature repositories to keep track of potentially harmful applications.

\begin{figure}[!t]
  \centering
    \includegraphics[scale=0.24,keepaspectratio]{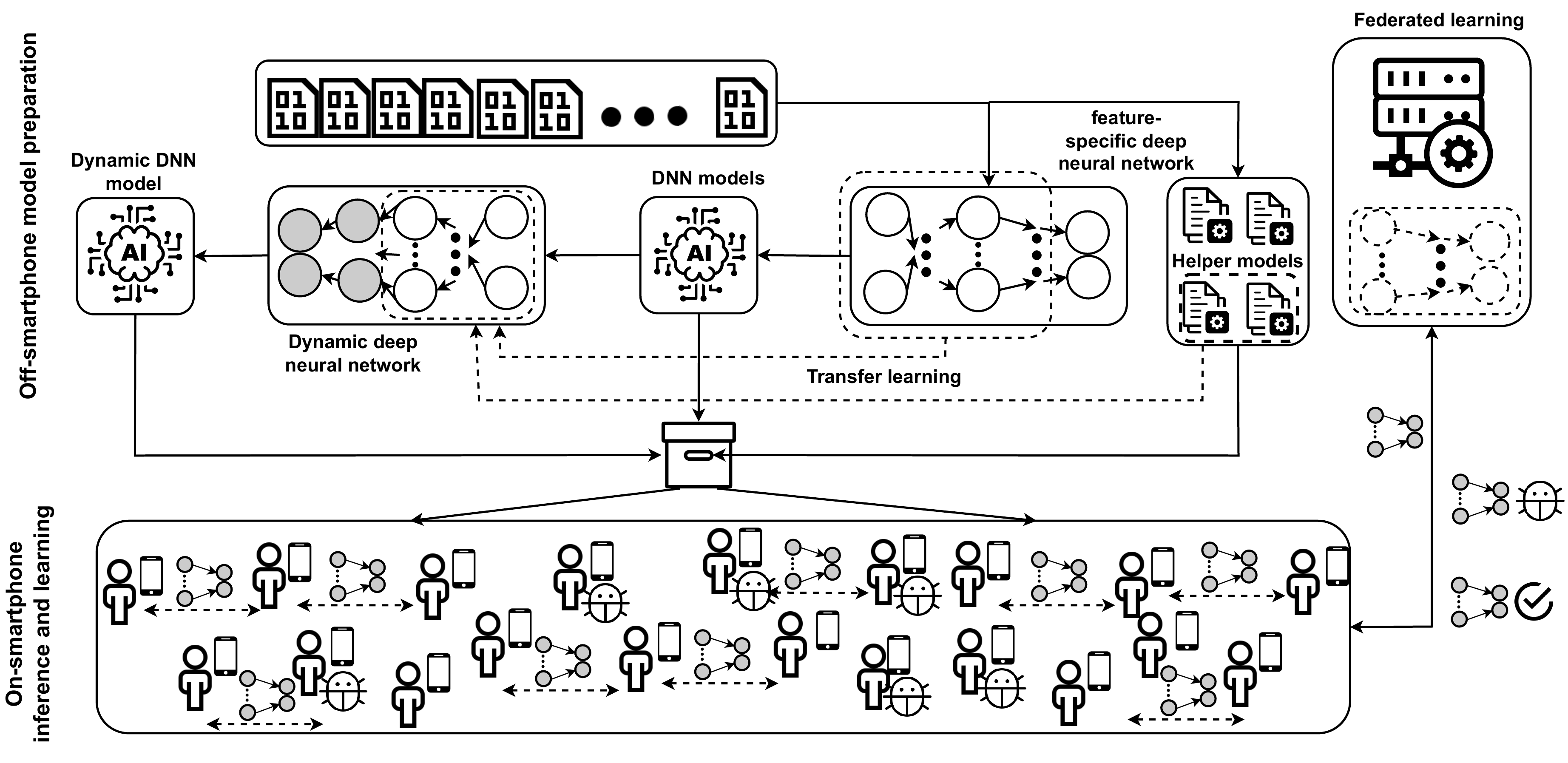}
 \caption{ADAM's overall structure consists of feature-specific DNN model preparation, including corresponding collaborative models for learning in the appearance of malicious participants. }
 \label{fig:model}
\end{figure}

This paper presents \textbf{A}nalytic-based \textbf{D}eep neural network \textbf{A}ndroid \textbf{M}alware detection \textbf{(ADAM)} that employs a fine-grained set of features referred to as \emph{application fingerprint} through in-depth analysis to train feature-specific DNNs. The overall structure of ADAM is shown in Figure \ref{fig:model}. ADAM employs feature-specific deep neural networks to have a consensus on labeling applications whose ground truth is unknown. Such consensus emerges from the primary DNN model named ``static'' which incorporates the application fingerprint and \emph{helper models} focusing on only a ``subset'' of features. These helper models are robust and representative and designed to mitigate the impact of wrongly labeled applications by the primary model, improving detection performance. For better adaptability of ADAM to new applications, ADAM adjusts itself across smartphones by using the transfer learning technique to recycle the pre-trained models and making them more \emph{universal} and \emph{collaborative} by model personalization and federated learning techniques. This adjustability is also protected by federated learning guards, which monitor ADAM to be immune to poisoning attacks. Federated learning model guards inherit the feature-specific and collaborative DNN structure to filter \emph{non-compliance} and \emph{malicious} participants from involving in the weight updates. ADAM  relies on a diverse dataset \cite{ml7,maldroid,androzoo} containing more than 153000 applications with over 41000 extracted features for training.

ADAM (available at https://github.com/amrmp/ADAM) is evaluated using real-world applications on different representative smartphones to showcase its performance regarding resource consumption and the accuracy of the DNN models. 

Results show that  the ADAM’s
feature-specific DNNs, on average, achieved more than 98\% accuracy. It results in outstanding detection performance being adjusted across smartphones for the collaborative model while robust against weight and feature (including the label) manipulation.

The rest of this paper is organized as follows: Section \ref{sec:relatedwork} discusses the background and reviews related work in malware detection, and Section \ref{sec:problem_statement} presents the proposed model fundamentals and methodology. 
We discuss experiments and results in Section \ref{sec:experiment} and conclude in Section \ref{sec:conclusion}.

\section{Background and Related Work}
\label{sec:relatedwork}

\subsection{Background}

\begin{figure}[!b]
  \centering
    \includegraphics[scale=0.4,keepaspectratio]{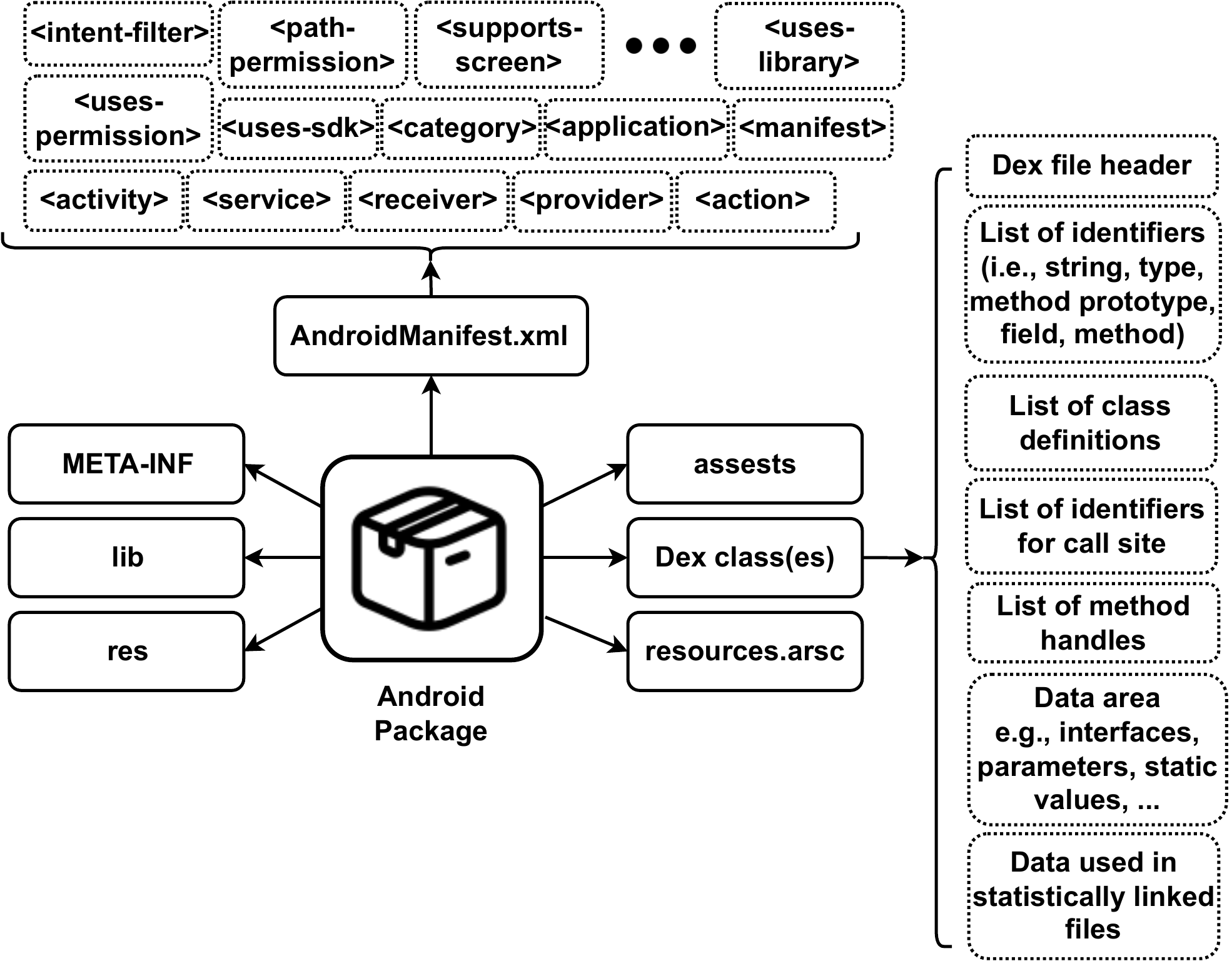}
 \caption{Android package content, and structure of essential parts of an Android application package; Android manifest file and Dex class(es).} 
 \label{fig:apk_structue}
\end{figure}

Google has developed the Android OS based on Linux OS, and since its emergence in 2008 by introducing Android 1.0, there have been many releases of this operating system. Many features for improving the OS's performance, security, and stability are added in each release, and some are depreciated. Applications for Android OS are written in Java/Kotlin, are packaged in the form of ``.apk'', and are compiled into Dalvik bytecodes (.dex). Each application package consists of ``AndroidManifest.xml", ``META-INF", ``lib", ``res", ``assests", ``resources.arsc", and ``Dex class(es)''. Figure \ref{fig:apk_structue} presents the overall structure of an Android package. 

In this structure, \emph{AndroidManifest.xml} provides essential information about the application and its components, including but not limited to activities, services, broadcast receivers, and content providers. Performing actions are based on \emph{intents} and the categories they belong to. For each application, a set of permissions and features is declared. The former explains required access to the underlying system. In the latter, the application may employ device-related features stated in the file. The folder \emph{META-INF} consists of the signature and the list of resources in the package. The folders \emph{lib} and \emph{res} present native libraries for running on specific device architectures (e.g., x86) and non-compiled resources such as images, respectively.
The \emph{assets} folder illustrates the raw source files provided by the developers with the application, and the \emph{resources.arsc} file describes compiled resources (e.g., strings) used in the application. Dex classes, on the other hand, present the application semantics meaning how the application logic is designed and converted to the code. More than one Dex class may exist in an application package, and these Dex classes hold information about the list of identifiers (e.g., strings or prototypes), class definitions, and method handles.

An extensive analysis of Dex classes leads to understanding the application's logical signature toward uncovering malware transformation techniques \cite{semanticanalysis}. Hence, ADAM looks for information regarding invoking methods and class APIs in the application Dex class(es). In particular, ADAM employs features with respect to invoking calls (i.e., implies for calling ``methods'' and may result in storing the values), referring to (1) invoke-virtual, which is a non-constructor and normal virtual method that is not private, static, or final, (2) invoke-super that is for invoking a closest superclass's virtual method with the same method restrictions, (3) invoke-direct that invokes a non-static direct method (i.e.,  a non-overridable instance method) which can be either a private instance method or a constructor, (4) invoke-static (i.e., a direct method), (5) invoke-interface for invoking interface method meaning whose definite class is unknown for an object, (6) invoke-custom for resolving and invoking the indicated call site into phases called call site resolution and call site invocation, and (7) invoke-polymorphic for invoking the indicated signature polymorphic method whose result may be stored with an appropriate move-result variant.

\subsection{Related Work}
Malware detection has been extensively studied from different perspectives (e.g., \cite{power1,power3,cloud2,ml2,cloud4,cloud5,network6}), in particular, usage of machine learning techniques \cite{ml5,ml7,ml10,ml12,ml13,ml14, ml16, analysis4,maldroid}.

Suarez-Tangil et al. \cite{power1} argued that externalized computation is the best energy-wise option by offloading specific functional tasks in a machine learning-based anomaly detection that relies on system calls. Caviglione et al. \cite{power3} presented a neural network and a decision tree-based technique for revealing covert communication (i.e., hidden data communication) usage by malware through monitoring energy consumption. 

Cui et al. \cite{cloud1} presented a service-oriented malware detection that captures and analyzes the network packets at an operator gateway assisted by clustering algorithms for analysis based on a knowledge directory. 
Zhou et al. \cite{ml2} inspected behaviors of applications by investigating the system calls and applied a machine learning classifier assisted by the Monte Carlo algorithm for adjusting the weights and the convergence. 
Tong and Yan \cite{cloud2} presented a hybrid approach in a modified Android OS and kernel that employs patterns of system calls related to the file and network access. A malicious pattern is generated based on the ratio of average frequency for sequential system calls in both malicious and benign sets. 

Sun et al. \cite{cloud4} argued that the core functions and behavior of malware within a malware family are similar; hence, they combined runtime behaviors with static structures for detecting malware variants. Hung et al. \cite{cloud5} presented a cloud-assisted malware detection for replicating the context of a mobile device in the cloud and on-device real-time mobile malware detector with the help of machine learning techniques (i.e., CNN and SVM). 

Android application internet activity or monitoring the network activity is studied in \cite{network6} through the use of graph theory techniques assisted by a security practitioner for characterizing malware samples. Zhang et al. \cite{analysis4} presented an n-gram based fingerprinting with the help of an online classifier. The risk assessment for permissions is studied in \cite{ml16} that emerges from unnecessary permissions given to Android applications with the help of the least square support vector machine. Analyzing the used permissions in the application manifest files can also be seen in \cite{ml10, ml10}. Inspecting the sequence of system calls is studied in \cite{maldroid} through a virtual machine introspection-based dynamic analysis system to extract the features and use deep learning for classification or considering common features shared in a malware family \cite{ml12}.

Potentially-harmful libraries across iOS and Android devices are studied in \cite{crossplat1}  to understand the behaviors of applications for both platforms for finding malicious behavior. Iterative feature selection for malware detection is discussed in \cite{ml5} through a lightweight classifier. 
Arp et al. \cite{ml7} presented a lightweight method for malware detection by employing static analysis of the applications (similar to \cite{ml14} for using Normalized Bernoulli Naive Bayes classifier) assisted by the SVM technique providing meaningful description.

\section{\textbf{A}nalytic-based \textbf{D}eep neural network \textbf{A}ndroid \textbf{M}alware detection \textbf{(ADAM)}}
\label{sec:problem_statement}

This section initially defines and formulates the malware detection problem and then explains the component of ADAM (Figure \ref{fig:model}) in detail.

For a set of smartphones ($\mathbb{D}$) with the size of $|\mathbb{D}|$, $\mathbb{F}^{i}_{j}$ denotes the \emph{application fingerprint} that is collected extensive feature set of the application $j$ from the smartphone ($D_i$) derived from parsing the manifest file and Dex class(es). 
The feature set ($\mathbb{F}^{i}_{j}$) consists of $n$ feature templates ($\mathbb{T} = \mathbb{T}_{1}\cup\cdot\cdot\cdot\cup\mathbb{T}_{n}$) represented as a one-hot vector with respect to the official Android developer guide, and belong to the analyzed components of the application $j$ shown in Figure \ref{fig:apk_structue}.

\begin{equation}
    \label{eq:feature_set1}
    \mathbb{F}^{i}_{j} = \bigcup_{k=1}^{n} \mathbb{T}_{k}.
\end{equation}

In Equation \ref{eq:feature_set1}, each $\mathbb{T}_k$ is an independent and identically distributed (IID) feature template set meaning $\forall \; \mathbb{T}_i \in \mathbb{T} : \mathbb{T}_{i}\cap\mathbb{T}_{j} = \emptyset, \; i \ne j $. 

For a given application $j$ and the corresponding feature set $\mathbb{F}^{i}_{j}$, ADAM employs the feature-specific DNN model $\Gamma = \Gamma_\psi\cup\Gamma_\omega\cup\Gamma_\eta$. This DNN model consists of the full feature set as static ($\Gamma_\psi$) and its corresponding collaborative ($\Gamma_\omega$) machine learning models. It includes different subsets of the features for representing the helper models (HMs) as $\Gamma_\eta$ to produce the classification probabilities of the application in the form of benign ($\mathbb{P}_\beta$) or malware ($\mathbb{P}_\mu$). These probabilities are obtained on smartphones and define the class of application $j$ that belongs to either benign $\beta$ or malware $\mu$.

\begin{equation}
\label{eq:probabilitya}
  \begin{aligned}
        \Gamma(\mathbb{F}^{i}_{j}) = \left\{
    \begin{array}{ll}
        \beta & \; \mathbb{P}_\beta \geq \mathbb{P}_\mu.\\
        \mu & \mathbb{P}_\beta < \mathbb{P}_\mu.
    \end{array}
\right. 
\end{aligned}
\end{equation}

In Equation \ref{eq:probabilitya}, $\Gamma$ is combination of the \emph{static}, \emph{collaborative}, and \emph{helper} DNN models that collaboratively contribute to $\Gamma(\mathbb{F}^{i}_{j})$.

\subsection{Feature-specific DNN Models}
\label{sec:model_generic}

\begin{figure}[!b]
  \centering
    \includegraphics[scale=0.31,keepaspectratio]{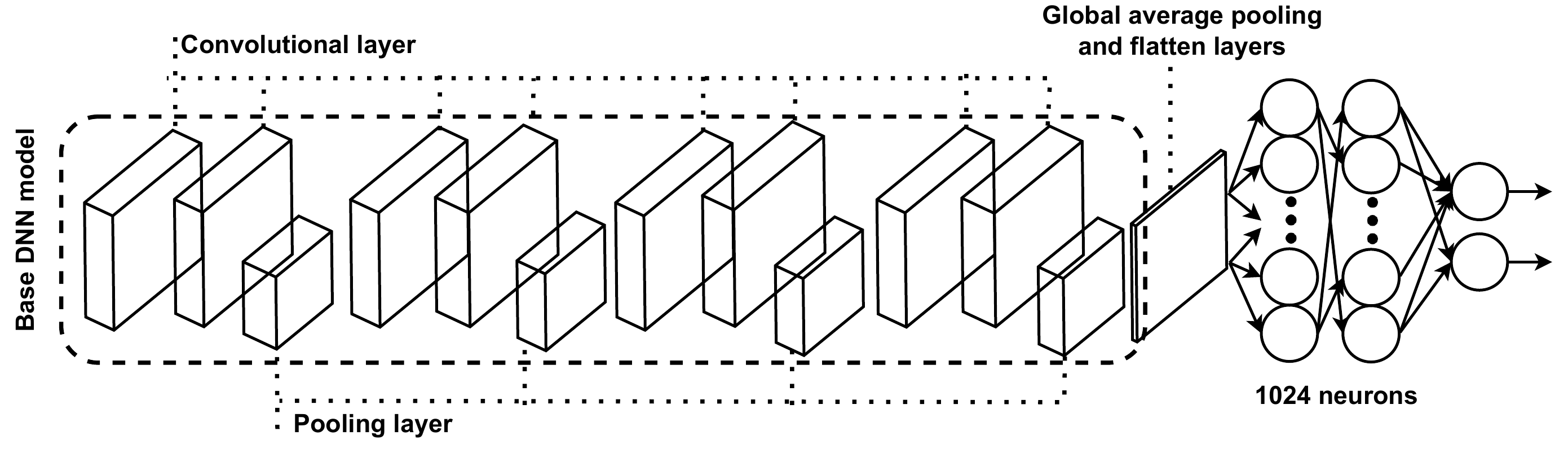}
 \caption{The structure of ADAM static DNN highlighting the \emph{base} part for transfer learning.} 
 \label{fig:dnn_static}
\end{figure}

\begin{figure}[!b]
  \centering
    \begin{minipage}[b]{1\columnwidth}
    \centering
    \includegraphics[scale=0.45,keepaspectratio]{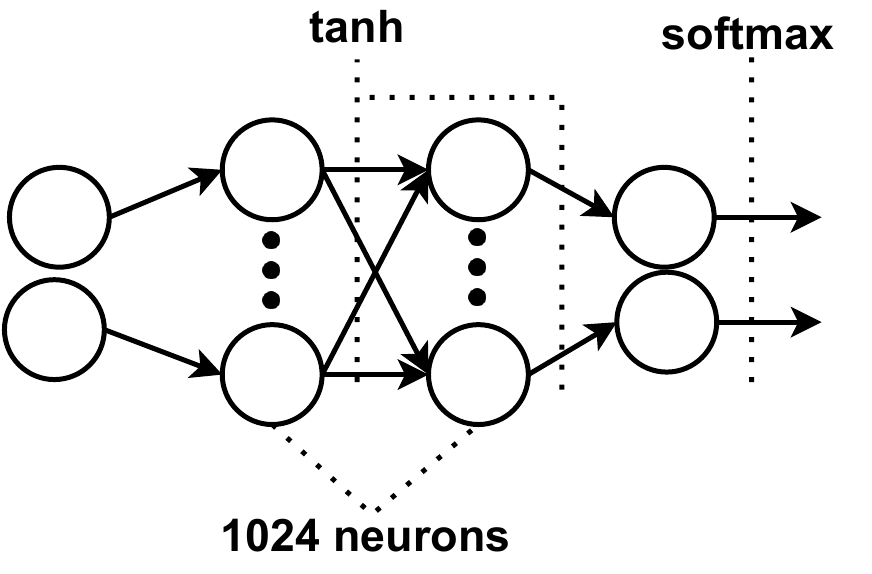}
    \subcaption{MLP-based helper models with two hidden layers.}
    \label{fig:helperespa}
  \end{minipage}
  \begin{minipage}[b]{1\columnwidth}
  \centering
    \includegraphics[scale=0.444,keepaspectratio]{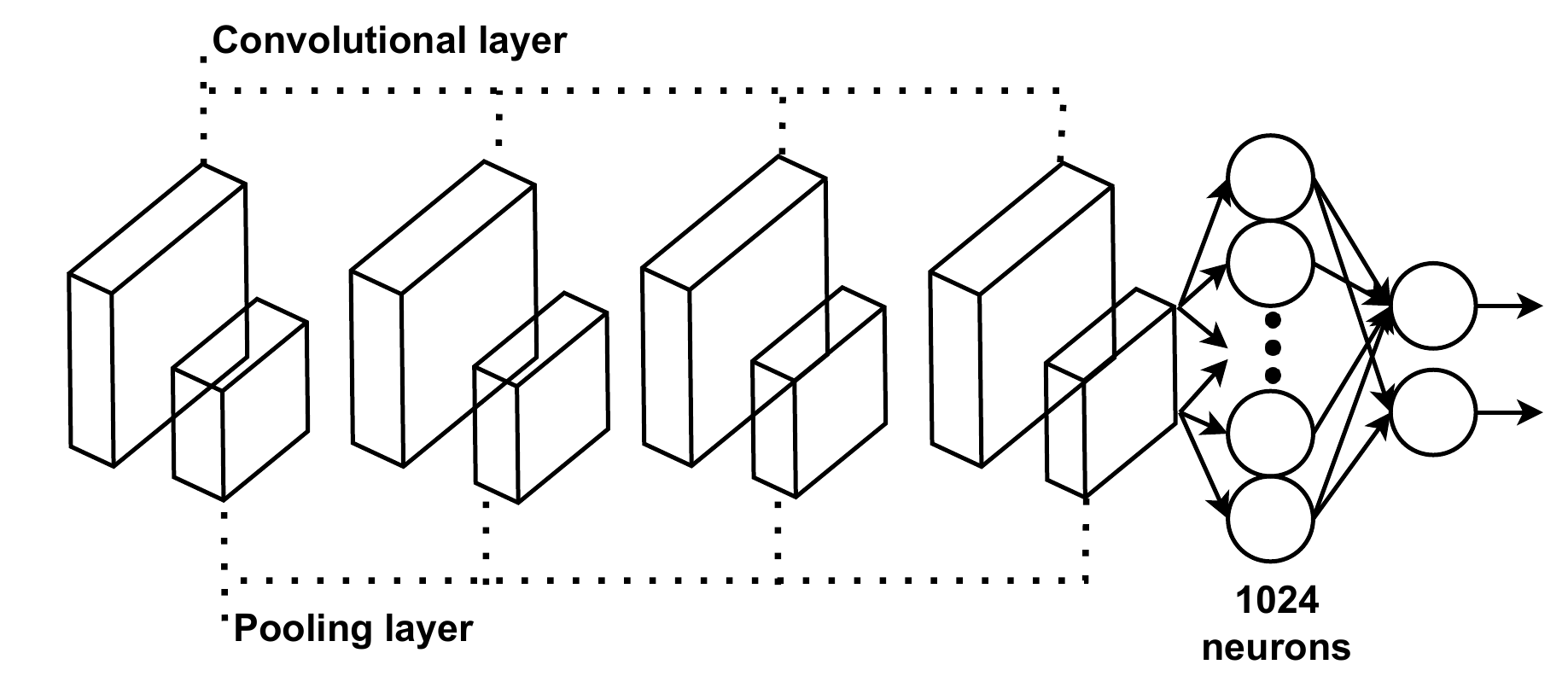}
    \subcaption{CNN-based helper models with four convolutional layers.}
    \label{fig:helperespb}
  \end{minipage}
    \caption{The overall structure of helper models for (a) multi-layer perceptron and (b) CNN neural networks.} 
    \label{fig:hlpersp}
\end{figure}

The \emph{aim} of feature-specific DNN models is to have a consensus on an application \emph{label} whose ground truth is unknown. These DNN models employ different classes of artificial neural networks (ANNs),  multi-layer perceptron (MLP) networks shown in Figure \ref{fig:helperespa} or convolutional neural networks (CNNs) in Figures \ref{fig:dnn_static} and \ref{fig:helperespb} based on the spatial correlation level in a (sub)set of features. The former is a classification neural network with less spatial correlation in the selected features. In the latter, increasing features may complicate the classification problem despite the existing correlation between Android APIs and the corresponding permissions/device features (e.g., using WiFi and Telephony API class). However, the semantics and logic behind such features might be left out; thus, they inherit the structure of the ADAM full application fingerprint (i.e., static model) while specifically designed for the target features. There basically exists inter-correlation between the sensitive APIs and their corresponding methods; however, this existence becomes more inter-connected with the relevant permissions due to different permission requirements for different elements of the manifest file.

Table \ref{tab:featurespm} represents the features used for DNN models; the full application fingerprint (i.e., the static model) and different \emph{subsets} of the fingerprint referred to as the helper models (HMs). This Table illustrates that the permission features\footnote{Permissions are obtained from each manifest tag that requires specific permissions (e.g., services, broadcast receivers, or activities), providing a comprehensive set of permissions that exists in the literature.} (including the corresponding protection levels) are the basis for the helper models. This is due to the correlation between permission features and other sensitive features, as their usability heavily depends on the granted permissions. Thus, these models cover the most well-studied features in the literature for malware detection and promote identifying which features may be beneficial for increasing performance detection. 

CNNs rely on multiple layers to extract high-level features from an input, in which each layer transforms the given input into a more abstract representation. Such representation evolves by passing through the layers where the initial ones focus more on extracting discriminate information than the successive layers involved in retrieving more semantic information. Layers of CNNs are fallen into feature learning and classification layers such that the former employs convolutional and pooling layers, including kernel filters, to perform feature extraction. The latter are fully connected layers and are recognized as regular neural networks with certain neurons. In contrast, MLPs are fully-connected feedforward neural networks that may consist of one or more hidden layers, where layers feed the successive layers with the result of the computation. 

A feature-specific DNN model accepts extracted features vector $X_f=[f_1, \dots, f_i]$ and the corresponding labels (i.e., classes), benign ($\beta$) or malware ($\mu$), in the form of $Y_f=[\beta, \mu]$. They produce  predicted output vectors in the form probabilities $Y_{f}^{'}=[\mathbb{P}_{\beta}, \mathbb{P}_{\mu}]$. The predicted output is used for training the network by calculating the loss values ($E_f$) when the trainable parameters (i.e., the CNN's weights) are modified based on the predicted output and the actual labels. Suppose $W_f$ represents feature-specific DNN model weights; the adjustment is conducted by $W_f \gets W_f - \gamma\nabla_{W_f}E_f$, where $\nabla$ is the vector differential operator and $\gamma$ is the learning parameter to manage the weight adjustment. The backpropagation algorithm controls the process that performs repetitive procedures in forwarding and backward passes for producing predicted outputs and adjusting the weights based on the computed error.

\begin{table}[t!]
    \centering
    \caption{Feature-specific DNN Models Strucutre}
    \begin{tabular}{|p{0.6cm}<{\centering}|p{0.8cm}<{\centering}|p{6cm}<{\centering}|}
    \hline
        Model & Category & Selected Features \\
    \hline
    Static & CNN & Android manifest attributes \cite{androidmanifest}, permissions, protection levels, features, intents, categories, providers, receivers, services, API classes, and API sensitive methods    \\ \hline
    HM1 & MLP & Permissions, protection levels, and features  \\ \hline
    HM2 & MLP & Permissions, protection levels, features, intents, categories, providers, receivers, and services \\ \hline
    HM3 & CNN & Permission, protection level, and features, API classes, and API sensitive methods  \\ \hline
    HM4 & MLP & Permission, protection level, and features, and API classes  \\ \hline
    HM5 & CNN & Permissions, protection levels, features, intents, categories, providers, receivers, services, and API classes \\ \hline
    HM6 & CNN & Permissions, protection levels, features, intents, categories, providers, receivers, services, API classes, and API sensitive methods\\ \hline
    \end{tabular}
    \label{tab:featurespm}
\end{table}

ADAM static DNN model uses a multi-layer CNN depicted in Figure \ref{fig:dnn_static}, which inherits a customized AlexNet CNN structure \cite{alexnet}. The static DNN model employs \emph{four} double-layer convolutional layers followed by a maximum pooling layer. While the structure of pooling layers is the same, having a pooling size of 2x2, each of the double-convolutional layers differs from the other, starting with the filter size of 16, followed by 32 and 64, and eventually 128 in the last double-convolutional layer. These convolutional layers utilize rectified linear activation function (ReLU), the kernel size of 3x3, and preserve spatial dimensions by setting the padding to ``same''. These layers are then connected to a global average pooling layer, followed by two fully connected layers consisting of 1024 neurons. The fully connected layers use a tangent hyperbolic ($tanh$) activation function, while the output layer employs a sigmoid activation function for delivering the results.

Feature-specific DNN models are used to produce a final label for a new and unseen smartphone application, which will later be used to prepare the collaborative model. This is done through semi-supervised learning by pseudo-label technique \cite{lee2013pseudo} that employs the maximum predicted probability of classes to set labels for unlabeled applications. Hence, the approximate labels based on the labeled data are examined, removing manual labeling of the unlabelled data. ADAM static DNN model (and associated model helpers) predicts two classes (i.e., benign or malware) with the corresponding probability of $\mathbb{P}_\beta$ and $\mathbb{P}_\mu$. For an unlabeled application $j$,  $l_j$ is the {\em pseudo-label} of the application, computed by $l_j = arg\max\{\Gamma_\psi,\Gamma_\eta\}$.

Even though the ADAM static DNN model \emph{likely} leads to providing an accurate pseudo label for the collaborative model, there might be situations a poor pseudo labeling would hurt the detection by falsely labeling the applications. Hence, helper models ($\Gamma_\eta$) are intended to lessen the effect of the static model's wrong pseudo-labeling. We empirically argue that in some cases, an application fingerprint, built by the extensive set of features, may become more \emph{concise} by selecting a subset of features for better classification. This is because the application developer might not fill in all the manifest files correctly or intentionally, focusing more on the most dominant elements of the application.

These helper models focus on the essential aspect of the collected application features (i.e., either the Dex classes or the manifest file) \cite{ml13,ml16,semanticanalysis}. They assist the static DNN model by providing probabilities of pseudo labels in the form of benign ($\mathbb{P}_\beta$) or malware ($\mathbb{P}_\mu$). In other words, for a given application $j$ on the device $i$ and the corresponding feature set $\mathbb{F}^{i}_{j}$, ADAM helper models employ ($\Gamma_\eta$) to provide a set of probabilities for the given applications.

\begin{equation}
    \label{eq:prob_set}
    \mathbb{P}_{i} = \bigcup_{k=1}^{|\eta|} {\Gamma}_{\eta}^{k} \cup{\Gamma}_{\psi}.
\end{equation}

In Equation \ref{eq:prob_set}, $\mathbb{P}_{i}$ represents the probability of the defined helper models, including the static model (${\Gamma}_{\psi}$), and $|\eta|$ represents the number of defined helper models. The set of probability ($\mathbb{P}_{i}$) is used to achieve a \emph{consensus} on the final pseudo label for the application.

The ADAM helper models obtain a set of essential features\footnote{They are extensively and comprehensively selected to the most widely used features in the literature.} chosen from declared feature templates and heir structures are shown in Figure \ref{fig:hlpersp}. The feature set ($\mathbb{F}^{i}_{j}$) consists of $n$ feature templates ($\mathbb{T} = \mathbb{T}_{1}\cup\cdot\cdot\cdot\cup\mathbb{T}_{n}$) presenting the application fingerprint. We define $\mathbb{T}_{\eth}$ in Equation \ref{eq:feature_sett1} as a subset of the most sensitive features chosen from the application fingerprint ($\mathbb{T}$).

\begin{equation}
    \label{eq:feature_sett1}
    \mathbb{T}_{\eth} = \{T_i,...,T_j\} \subset \mathbb{T}, \; 1 \leq |\mathbb{T}_{\eth}|< |\mathbb{T}|.
\end{equation}

The ADAM helper models follow the Boyer-Moore voting method \cite{boyer1991mjrty} as the consensus mechanism to label an application whose ground truth cannot be decided. It intends to determine the majority label among inferred labels produced by the DNN models with more than $N/2$ occurrences.

\subsection{Collaborative DNN Model}
\label{sec:model_personal}

Feature-specific DNN models, particularly the static and CNN-based helper models, exploit massive datasets for training, resulting in \emph{rich and pre-trained} DNN models. Such pre-trained DNN models can be recycled for a similar problem by retaining the mostly feature learning layers (i.e., learning parameters). This enables its adaptability for a universal DNN model by attaching a new set of classification layers. This functionality is assisted by transfer learning and involves splitting the pre-trained model into two parts, referred to as \emph{base} and \emph{head}. The base region inherits the weights of the pre-trained model, which are untrainable and unchangeable, attached to a new neural network with the target classes for the problem. This new neural network is the head of the new DNN model trained based on new data provided for the problem. 

\begin{figure}[!t]
  \centering
    \includegraphics[scale=0.46,keepaspectratio]{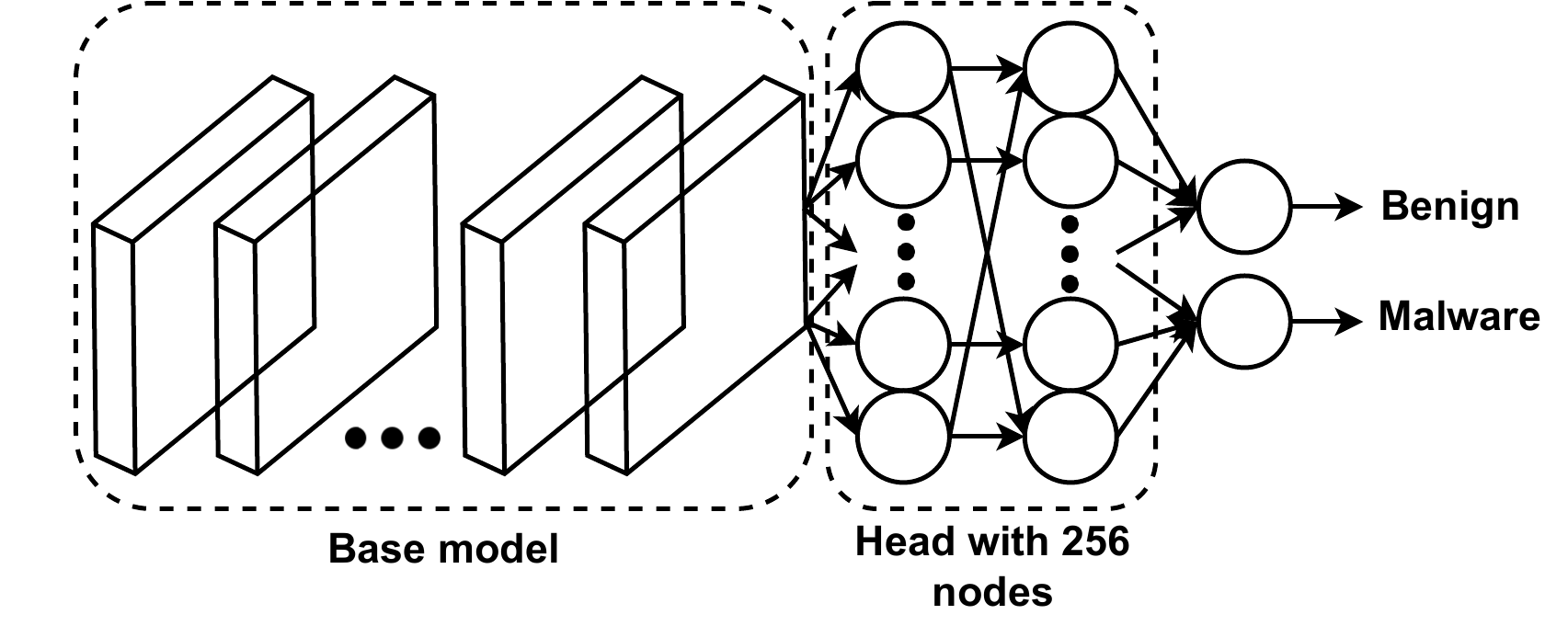}
 \caption{Two fully connected dense layers construct the head's region of ADAM collaborative DNN.}
 \label{fig:dnn_dynamic}
\end{figure}

For malware detection, transfer learning benefits ADAM, particularly by creating a universal and collaborative model. In Figure \ref{fig:dnn_static}, the base part is attached to a new neural network shown in Figure \ref{fig:dnn_dynamic} and is to be used for training and inference on the device. The structure of the new neural network is application-depended and should necessarily consider what type of resources are utilized for training. Therefore, it should be lightweight for the malware detection problem on smartphones, implying the training should consider their limited resources. In addition, such a neural network trained on smartphones takes advantage of preserving privacy and keeping data on devices. 

The collaborative DNN model employs the inferred pseudo-labels to train itself in a semi-supervised way.

In the pseudo-label technique, the overall loss $L$ for the collaborative DNN model is controlled by a weighted loss function, in which $\delta$ \ controls the contribution of unlabelled data \cite{lee2013pseudo}. Increasing the $\delta$ throughout training will result in more contribution from the pseudo-labeled data than initially focusing more on the labeled data. Thus, if $L_{\Lambda}$ represents the labeled data loss function and $L_{\mathbb{E}}$ shows the loss function of unlabelled data, the total loss is computed as $L = L_{\Lambda} + \delta L_{\mathbb{E}}$. The loss value for either $L_{\Lambda}$ or $L_{\mathbb{E}}$ is computed as $\frac{1}{n}\sum_{m=1}^{n}\sum_{i=1}^{2}L(x_i^m,y_i^m)$,  where the tuple ($x_i,y_i$) is the input sample and its corresponding output with a total sample of $n$.

Although the pseudo-label technique benefits ADAM, ADAM still lacks improving the trainable parameters across smartphones since each smartphone centralizes the training data and trainable parameters. Hence, federated learning enables smartphones to share trainable parameters with others in the network.

The federated learning technique employs an iterative process (i.e., rounds ($r_k$)) where the ADAM collaborative model state after local training is transmitted to a server, carrying the potential portion of model weight updates residing on the smartphone in each round. It will be aggregated into a single ADAM collaborative model state, containing the whole or partial model weights from smartphones. This update can be synchronous, where each smartphone will receive a model weight state of all the connected smartphones. The model weight state, however, can also be asynchronous by incorporating model updates from a portion of smartphones involved in a round and can happen in the presence of different smartphones' computational capacities or network delays. 

In a network, $n$ smartphones ($s_i$) are connected to a federated learning server. Each networked smartphone holds $t$ samples for training the ADAM collaborative DNN model ($f_i^s$). These samples account for a portion of the collaborative DNN model training parameters known as weights shared across the networked smartphones. We define $F^r_k$ that represents the aggregated ADAM collaborative DNN model weights at the round $r_k$.

\begin{equation}
\label{eq:fl}
F^r_k = \bigcup_{i=1}^{n} f_i^s, \;  f_i^s= \bigcup_{j=1}^{t} {\Gamma}_{\omega}(x_j^i,y_j^i).
\end{equation}

$F^r_k$ in Equation \ref{eq:fl} is, however, prone to data poisoning attacks (e.g., weight manipulation or label-flipping attacks), which consequently results in the poor performance of the model. One or a set of malicious participants manipulates the training parameters in each round of aggregation to harm the collaborative model.

ADAM employs an evaluation-based approach at the federated learning server to mitigate the impact of data poisoning attacks. ADAM builds the collaborative model on the server side, where it inherits the structure of the model on smartphones, freezing the base weights and adding new classification layers for training. The new classification layers resemble on-device training and aim to provide a non-poisoned but robust evaluator to mitigate attacks. Such models are trained with a set of vendor's stock ROM (i.e., all the operating system applications), in which applications on such phones, due to the integrity of the Android operating system, are considered benign. Such applications provide a set of diverse and effective evaluation test data and a specific performance on the particular model to identify malicious participants during model updating.

\section{Evaluation}
\label{sec:experiment}

This section explains the details of feature-specific DNN model training, transfer learning, and collaborative learning. 
We then present experimental results regarding model accuracy, resource consumption, and communication latency. ADAM's performance is also evaluated for model poisoning attacks, such as weight and feature manipulation, including label flipping attacks. 

\subsection{Feature-Specific DNN Model Training}

\begin{figure}[!t]
  \centering
    \begin{minipage}[b]{0.49\columnwidth}
    \centering
    \includegraphics[scale=0.41,keepaspectratio]{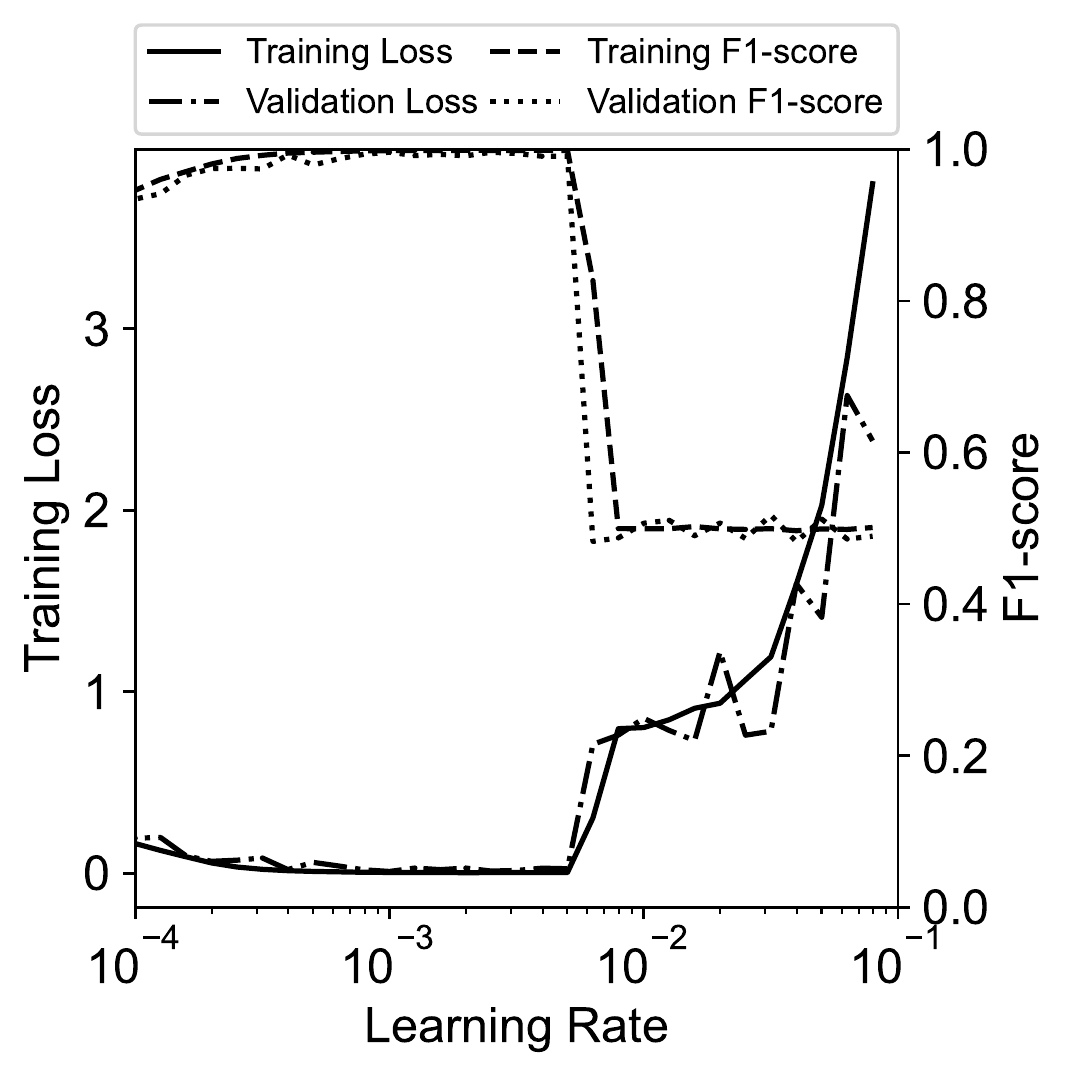}
    \subcaption{Loss}
    \label{fig:lrloss}
  \end{minipage}
  \begin{minipage}[b]{0.49\columnwidth}
  \centering
    \includegraphics[scale=0.41,keepaspectratio]{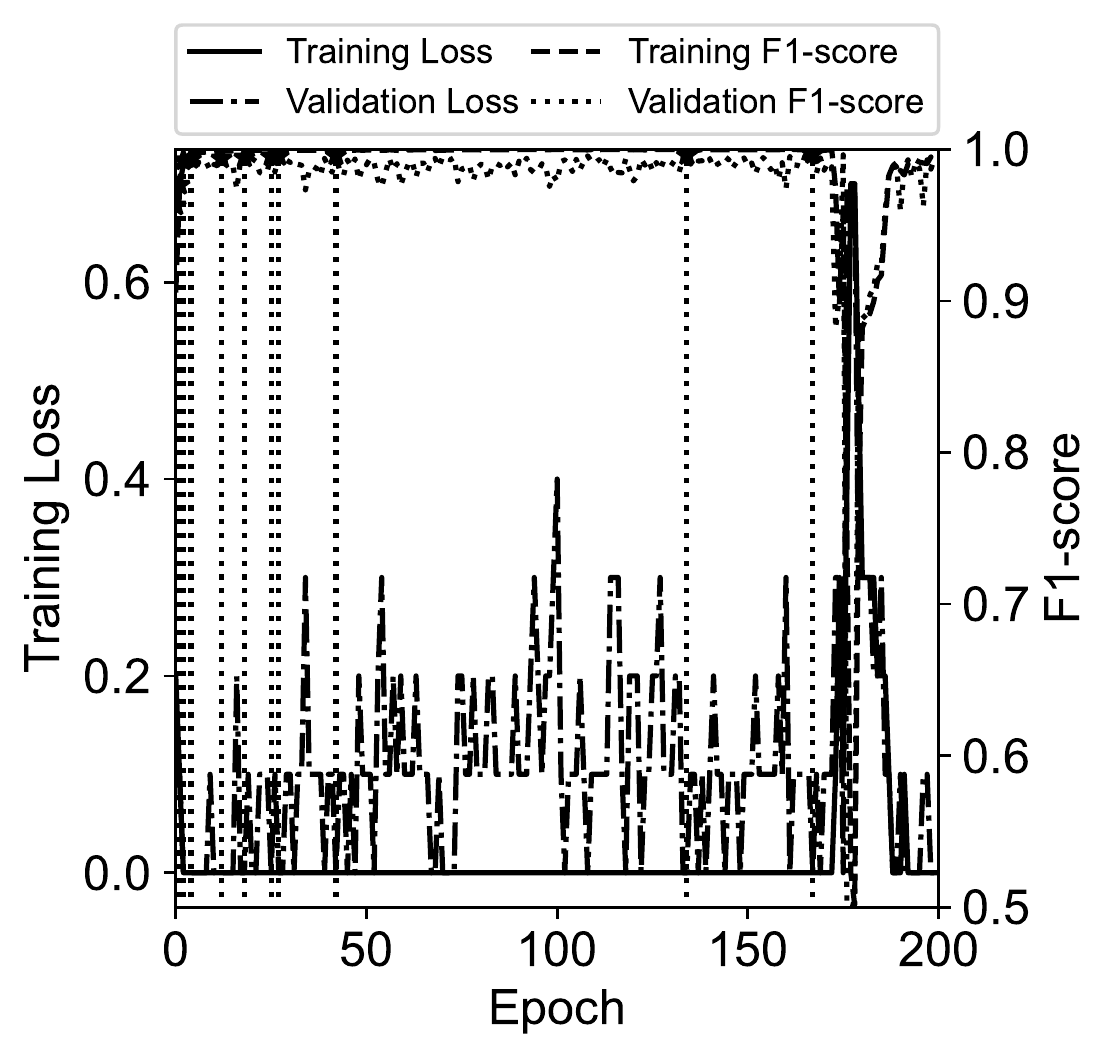}
    \subcaption{Accuracy}
    \label{fig:lraccu}
  \end{minipage}

 \caption{ADAM's static DNN performance (loss and f1-score) under different learning rates and the optimal value of $\sim$0.001. The vertical lines present the validation accuracy improvement throughout training.}
 \label{fig:learningrate}
\end{figure}

The ADAM employs real-world applications from a diverse range of applications in terms of size, application category, and release date in the form \emph{malware} and \emph{benign}. This dataset \cite{pasdar2022maps} contains collected applications from \cite{ml7,maldroid,androzoo} shown in Table \ref{tab:dataset_info}. The prepared dataset is divided into three parts with the ratio of $\sim$80:10:10 for training, validation, and testing datasets, respectively.

\begin{table}[h!]
    \centering
    \caption{Dataset Information}
    \begin{tabular}{|p{1cm}<{\centering}|p{1.2cm}<{\centering}|p{1.5cm}<{\centering}|p{1.5cm}<{\centering}|p{1.5cm}<{\centering}|}
    \hline
        Category & \#Total APKs & \#Processed Training APKs & \#Processed Validation and Test APKs$\dagger$ & \#Non-processed APKs \\
    \hline
    Benign & 76226 & 61350 & 14463 & 413 \\ \hline
    Malware & 76993 & 61350 & 13852 & 1791 \\ \hline
    \multicolumn{5}{c}{$^\dagger$The ratio for validation and test datasets is 50:50.}\\ 
    \end{tabular}
    \label{tab:dataset_info}
\end{table}

The feature-specific DNN models are implemented in Keras API trained on AWS accelerated EC2 instance \emph{g4dn.xlarge} that benefits from NVIDIA T4 Tensor Core GPUs. The performance evaluation is based on the training and validation loss and the corresponding accuracy in terms of {\em f1-score}. The f1-score presents the ratio of correct positive results divided by the number of all relevant samples, and the higher f1-score points out, the better the DNN model is trained.

The first part of training is determining the optimal learning rate for the static DNN structure. Hence, the performance of the ADAM static DNN model is evaluated under learning rates chosen from [$10^{-4}$,$10^{-1}$] with the batch size \emph{empirically} set to 32\footnote{Hyperparameter optimization based on different batch sizes led to different results. Larger batch sizes (i.e., $>32$) would fail due to out-of-memory (OOM) runtime errors. Smaller batch sizes (i.e.,  $\leq16$) could not provide the highest training performance in loss and f1-score but would increase the training time.}. The optimal value is set for training the ADAM static DNN model over 200 epochs in the next step. The chosen epoch value is evaluated over training for a range of [100-200] due to model improvements in this domain. The implemented DNN on Keras API leverages the ``checkpointing'' feature that monitors validation accuracy for saving the best-trained DNN model. 

Figure \ref{fig:learningrate} presents the ADAM static DNN performance for different learning rates. Figure \ref{fig:lrloss} illustrates that the learning rate close to $10^{-3}$ achieved higher training and validation accuracy shown in Figure \ref{fig:lraccu}. Hence, this optimal learning rate is used for training the ADAM static DNN model, and performance metrics are shown in Figure \ref{fig:lraccu}. The overall training time is nearly three days. The static DNN model achieved the lowest loss value for the training throughout training. Since the Keras API script monitors the validation accuracy and saves the better models, the ADAM static DNN model achieved an accuracy of 99.46\% for the test dataset. The vertical lines in Figure \ref{fig:lraccu} illustrate the checkpoints where the ADAM static model improved the validation accuracy for training.

\begin{figure}[!t]
  \centering
    \begin{minipage}[b]{0.49\columnwidth}
    \centering
    \includegraphics[scale=0.41,keepaspectratio]{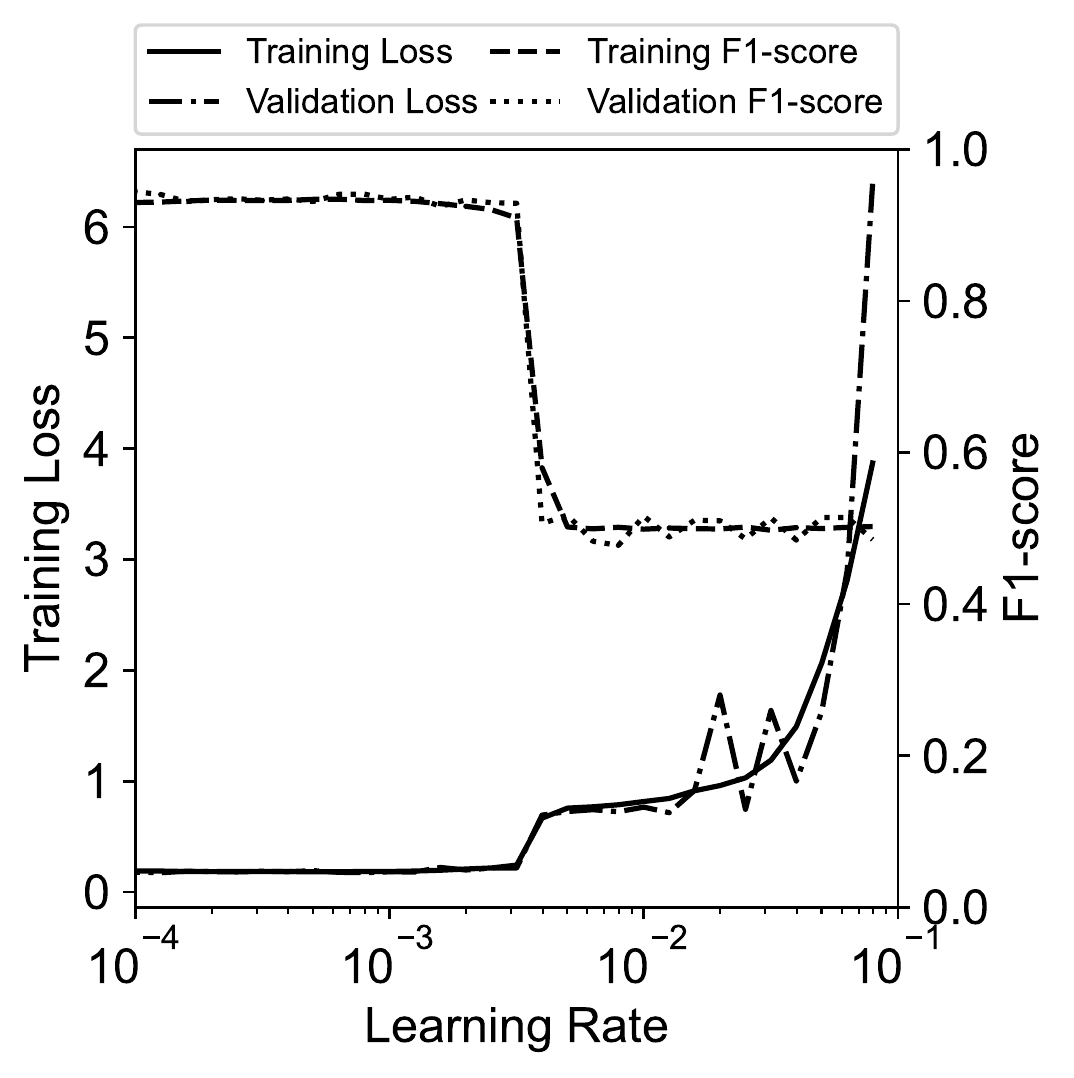}
    \subcaption{Loss}
    \label{fig:m1lrloss}
  \end{minipage}
  \begin{minipage}[b]{0.49\columnwidth}
  \centering
    \includegraphics[scale=0.41,keepaspectratio]{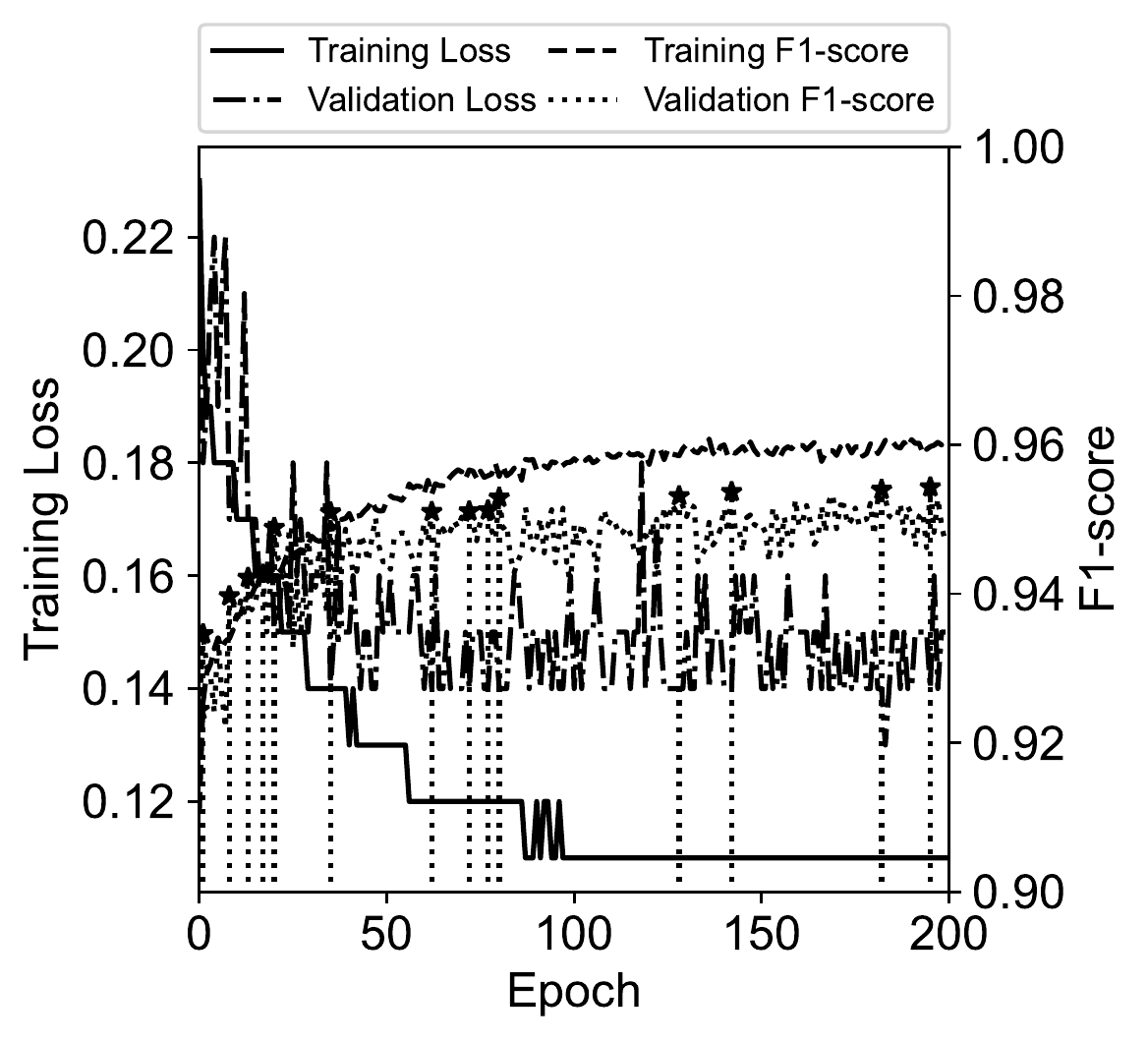}
    \subcaption{Accuracy}
    \label{fig:m1lraccu}
  \end{minipage}

 \caption{HM1 DNN performance (loss and f1-score) under different learning rates and the optimal value of $\sim$0.0001 with the accuracy of 95.86\%.
 }
 \label{fig:m1}
\end{figure}

\begin{figure}[!t]
  \centering
    \begin{minipage}[b]{0.49\columnwidth}
    \centering
    \includegraphics[scale=0.41,keepaspectratio]{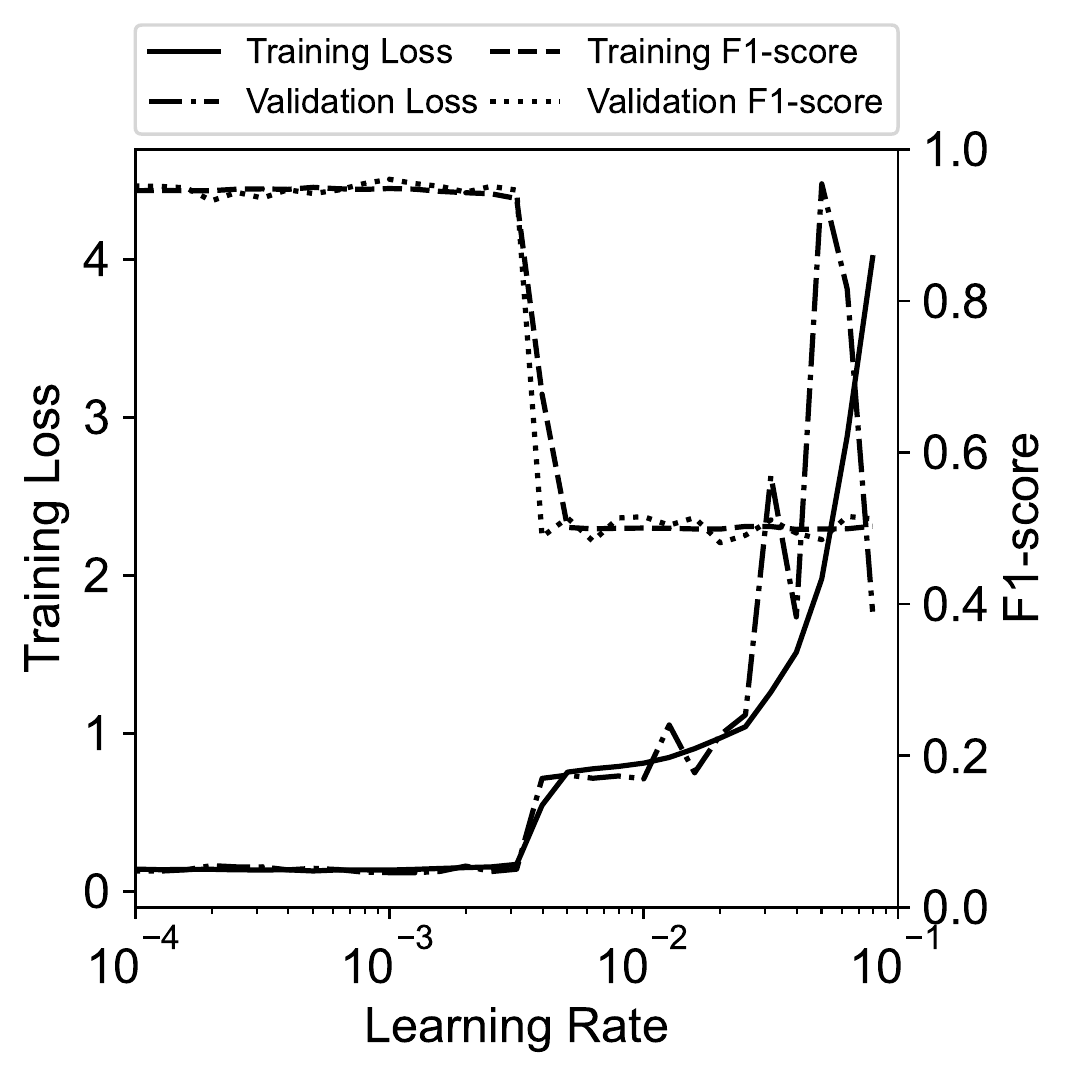}
    \subcaption{Loss}
    \label{fig:m2lrloss}
  \end{minipage}
  \begin{minipage}[b]{0.49\columnwidth}
  \centering
    \includegraphics[scale=0.41,keepaspectratio]{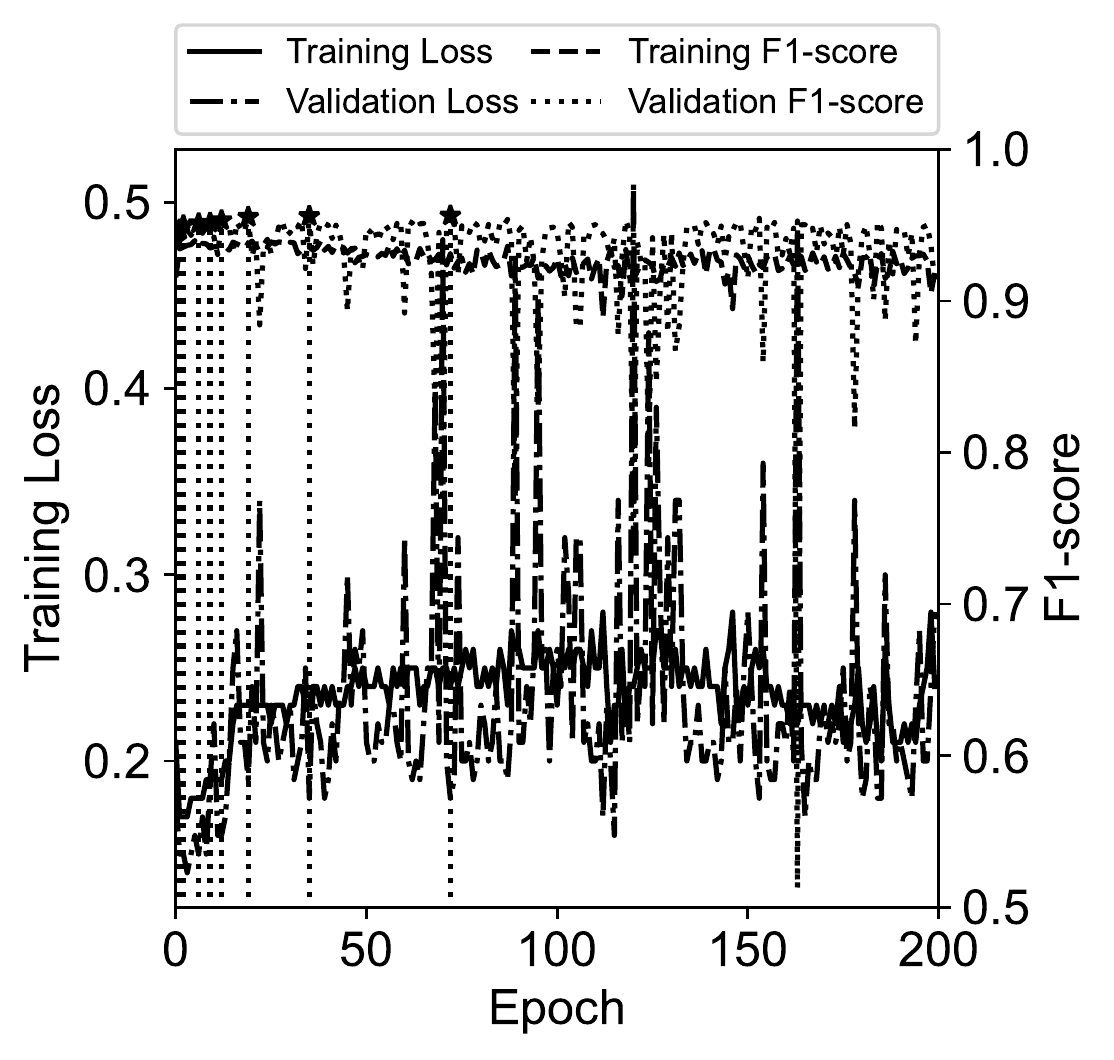}
    \subcaption{Accuracy}
    \label{fig:m2lraccu}
  \end{minipage}

 \caption{HM2 DNN performance (loss and f1-score) under different learning rates and the optimal value of $\sim$0.001 with the accuracy of 95.39\%.
 }
 \label{fig:m2}
\end{figure}

\begin{figure}[!t]
  \centering
    \begin{minipage}[b]{0.49\columnwidth}
    \centering
    \includegraphics[scale=0.4,keepaspectratio]{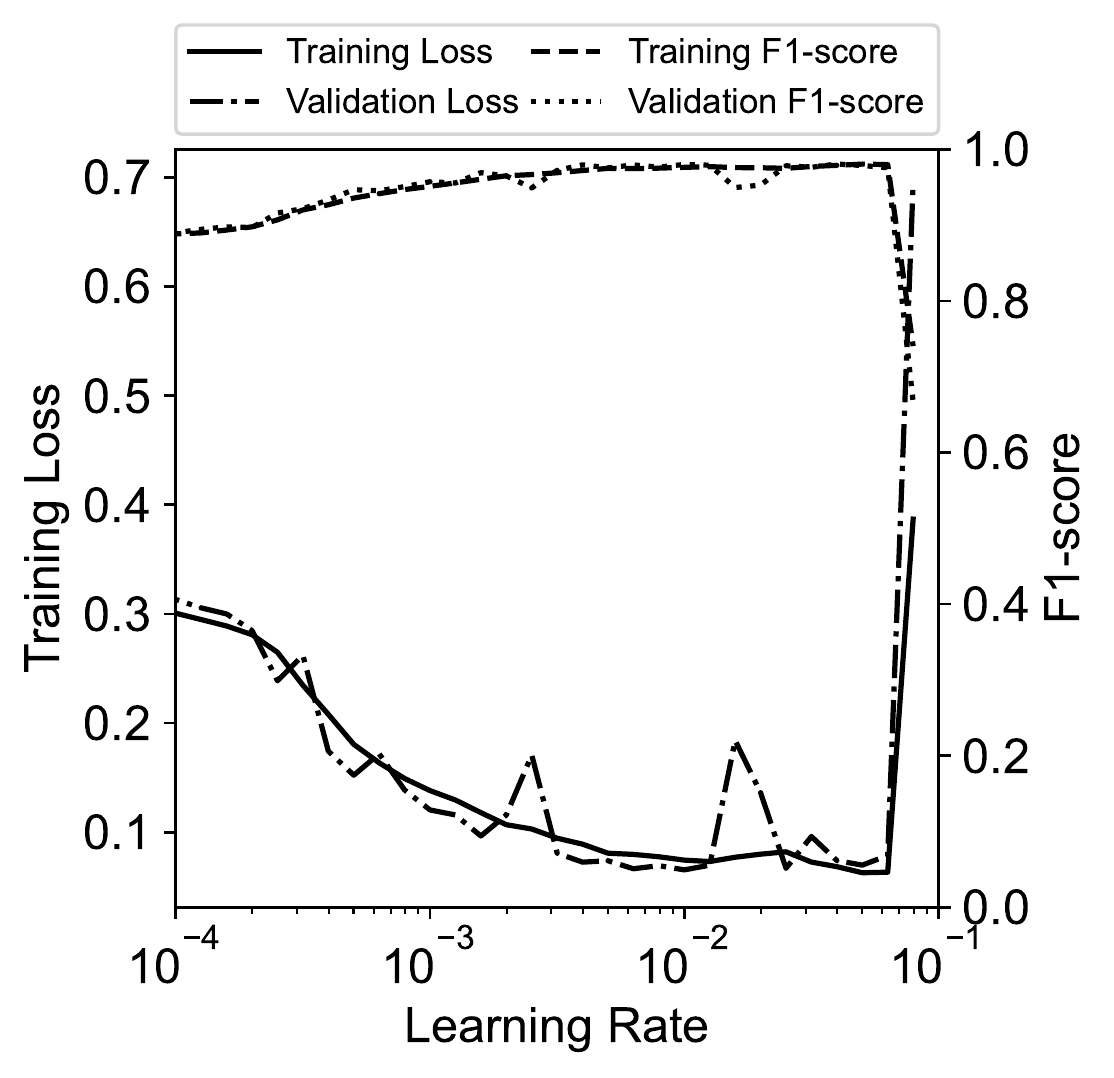}
    \subcaption{Loss}
    \label{fig:m3lrloss}
  \end{minipage}
  \begin{minipage}[b]{0.49\columnwidth}
  \centering
    \includegraphics[scale=0.4,keepaspectratio]{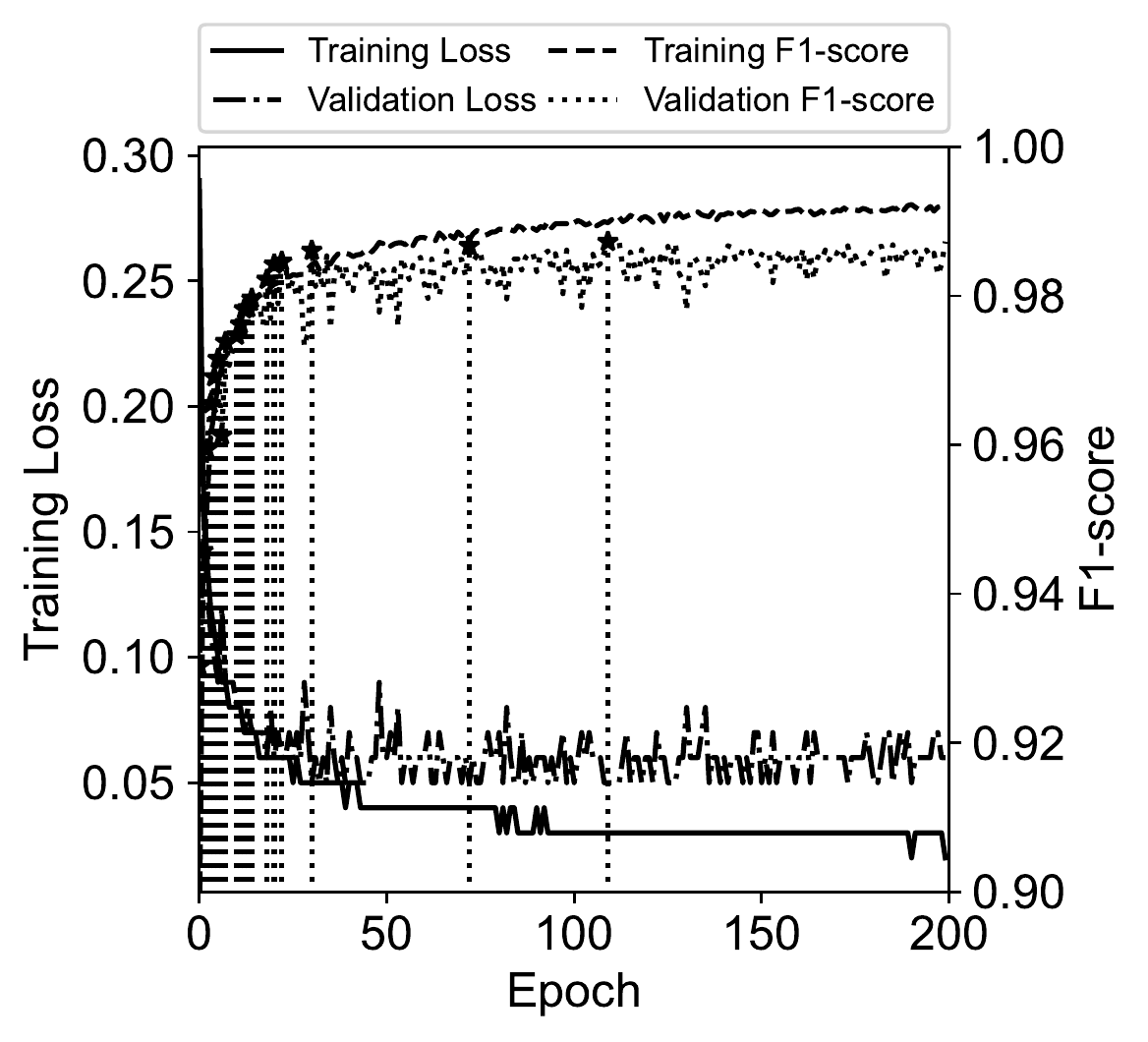}
    \subcaption{Accuracy}
    \label{fig:m3lraccu}
  \end{minipage}

 \caption{HM3 DNN performance (loss and f1-score) under different learning rates and the optimal value of $\sim$0.00398 with the accuracy of 98.88\%.
 }
 \label{fig:m3}
\end{figure}

\begin{figure}[!t]
  \centering
    \begin{minipage}[b]{0.49\columnwidth}
    \centering
    \includegraphics[scale=0.41,keepaspectratio]{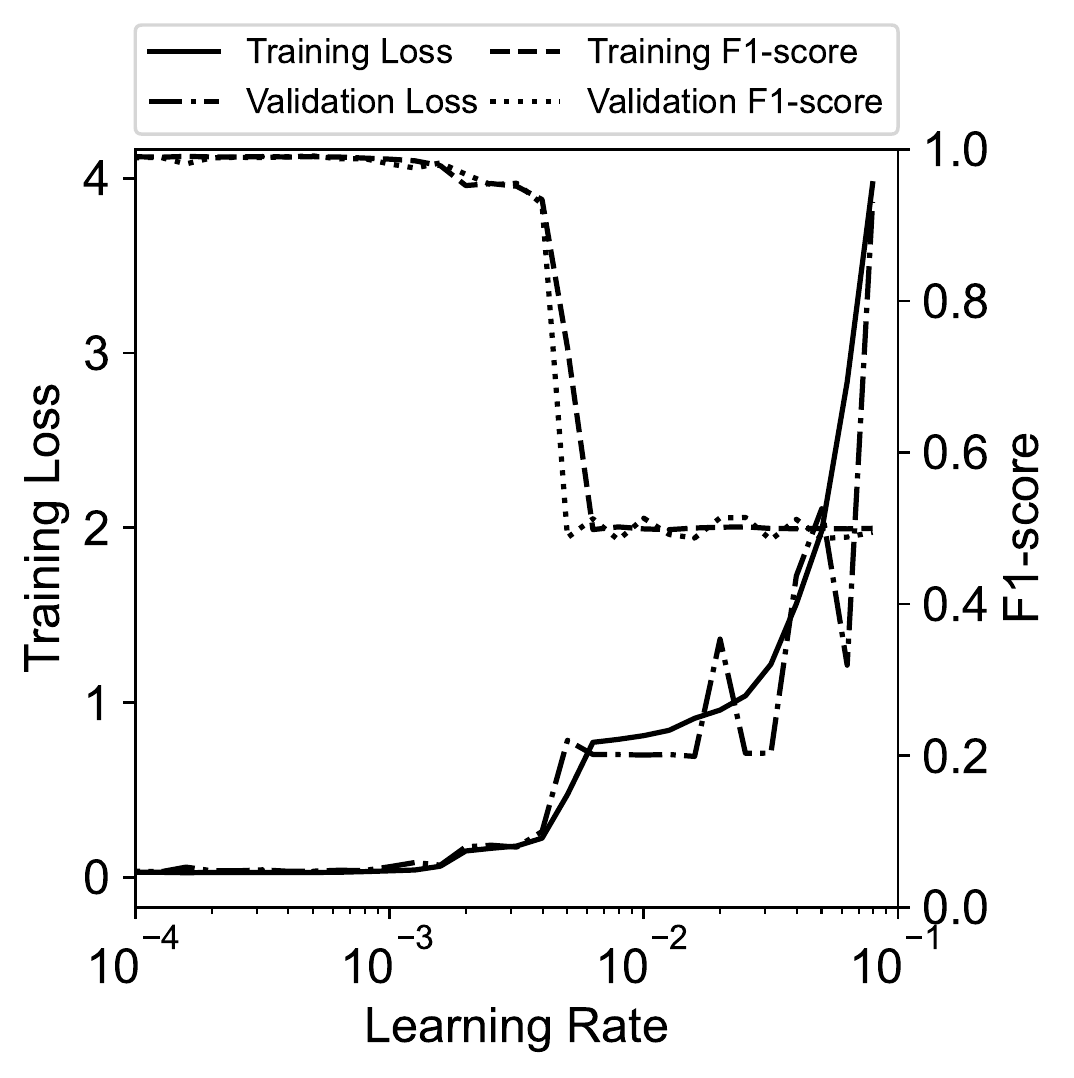}
    \subcaption{Loss}
    \label{fig:m4lrloss}
  \end{minipage}
  \begin{minipage}[b]{0.49\columnwidth}
  \centering
    \includegraphics[scale=0.41,keepaspectratio]{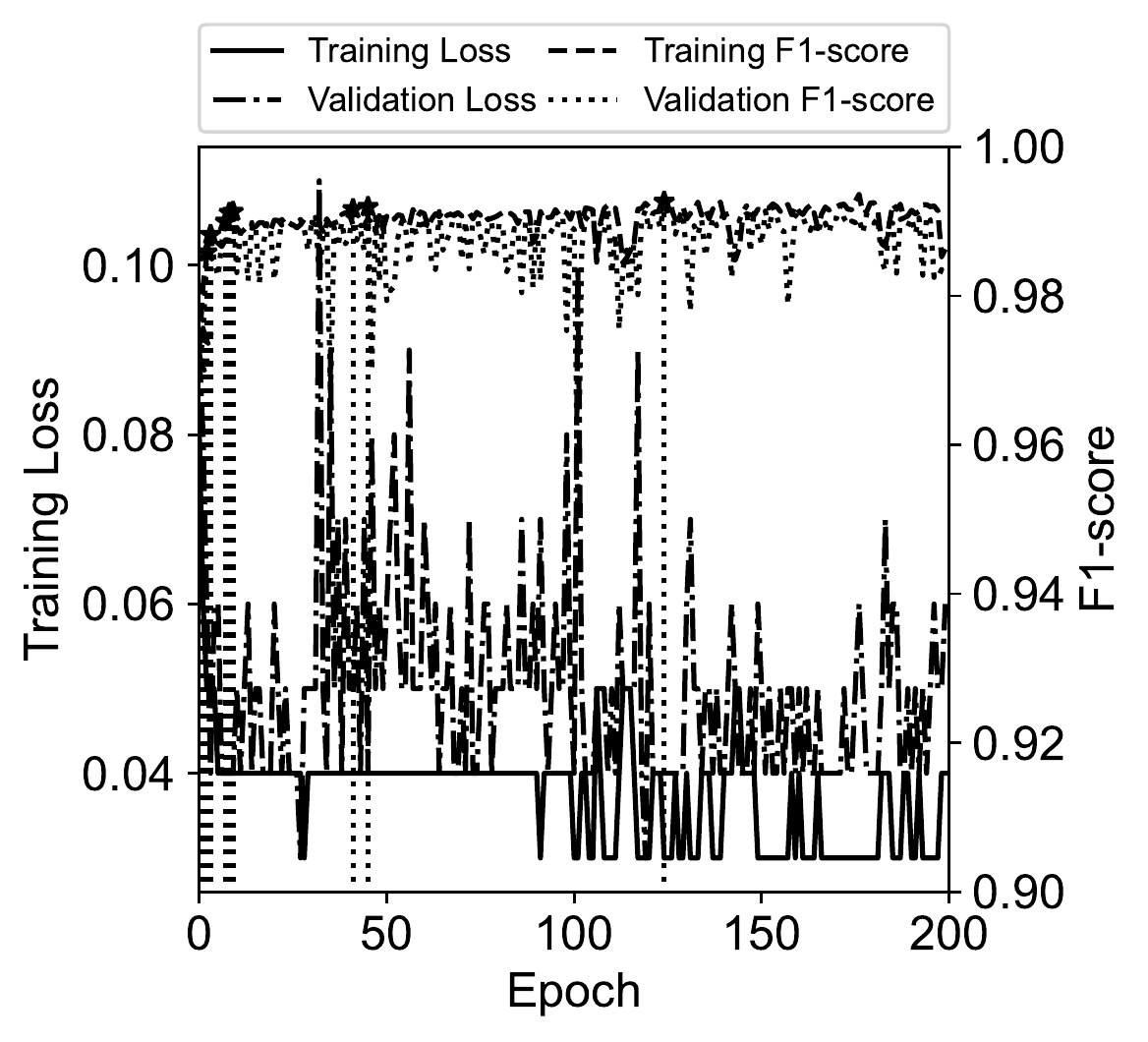}
    \subcaption{Accuracy}
    \label{fig:m4lraccu}
  \end{minipage}

 \caption{HM4 DNN performance (loss and f1-score) under different learning rates and the optimal value of $\sim$0.0005 with the accuracy of 99.23\%.
 }
 \label{fig:m4}
\end{figure}

\begin{figure}[!t]
  \centering
    \begin{minipage}[b]{0.49\columnwidth}
    \centering
    \includegraphics[scale=0.4,keepaspectratio]{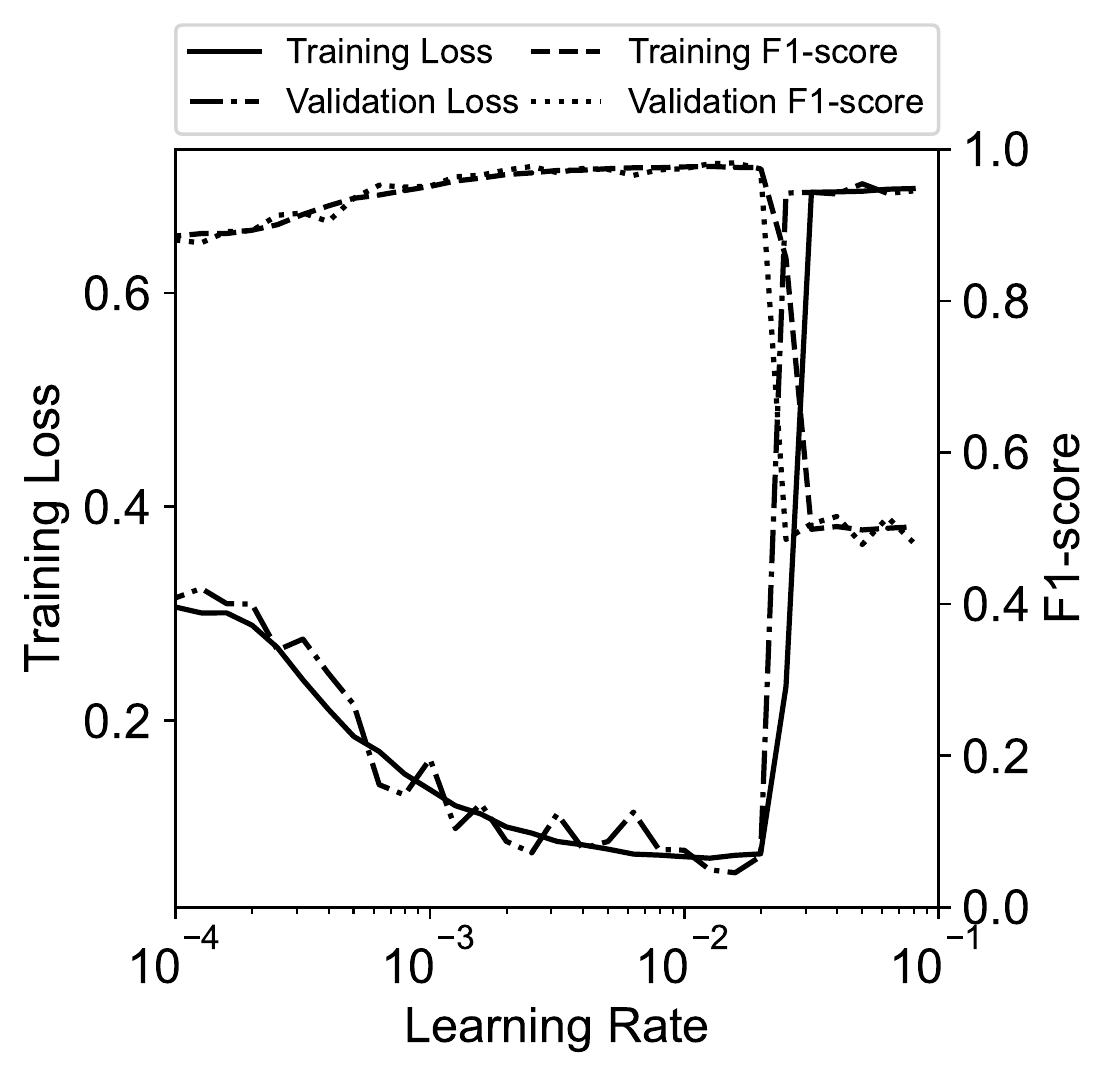}
    \subcaption{Loss}
    \label{fig:m5lrloss}
  \end{minipage}
  \begin{minipage}[b]{0.49\columnwidth}
  \centering
    \includegraphics[scale=0.4,keepaspectratio]{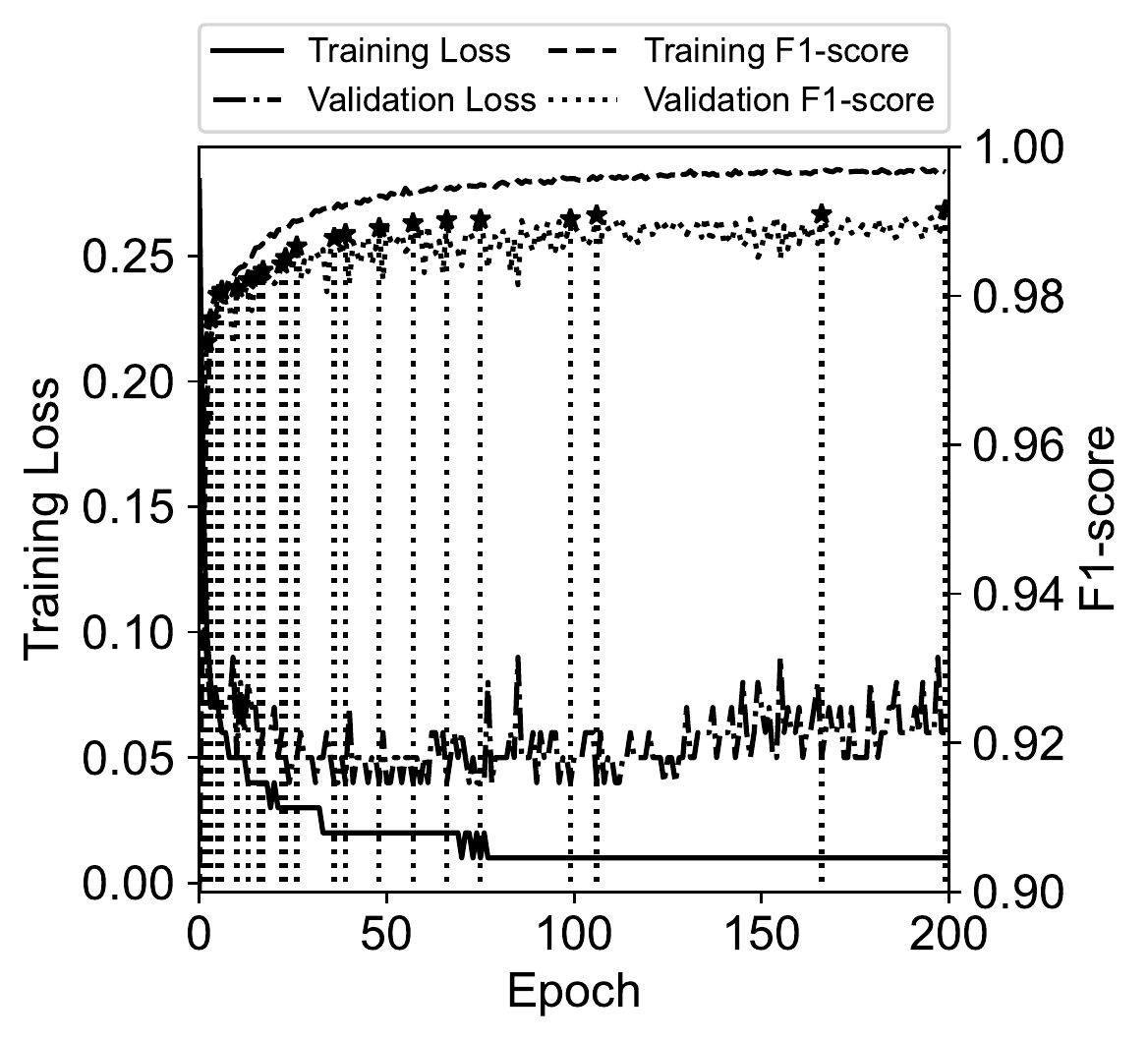}
    \subcaption{Accuracy}
    \label{fig:m5lraccu}
  \end{minipage}

 \caption{HM5 DNN performance (loss and f1-score) under different learning rates and the optimal value of $\sim$0.002511 with the accuracy of 98.99\%.
 }
 \label{fig:m5}
\end{figure}

\begin{figure}[!t]
  \centering
    \begin{minipage}[b]{0.49\columnwidth}
    \centering
    \includegraphics[scale=0.40,keepaspectratio]{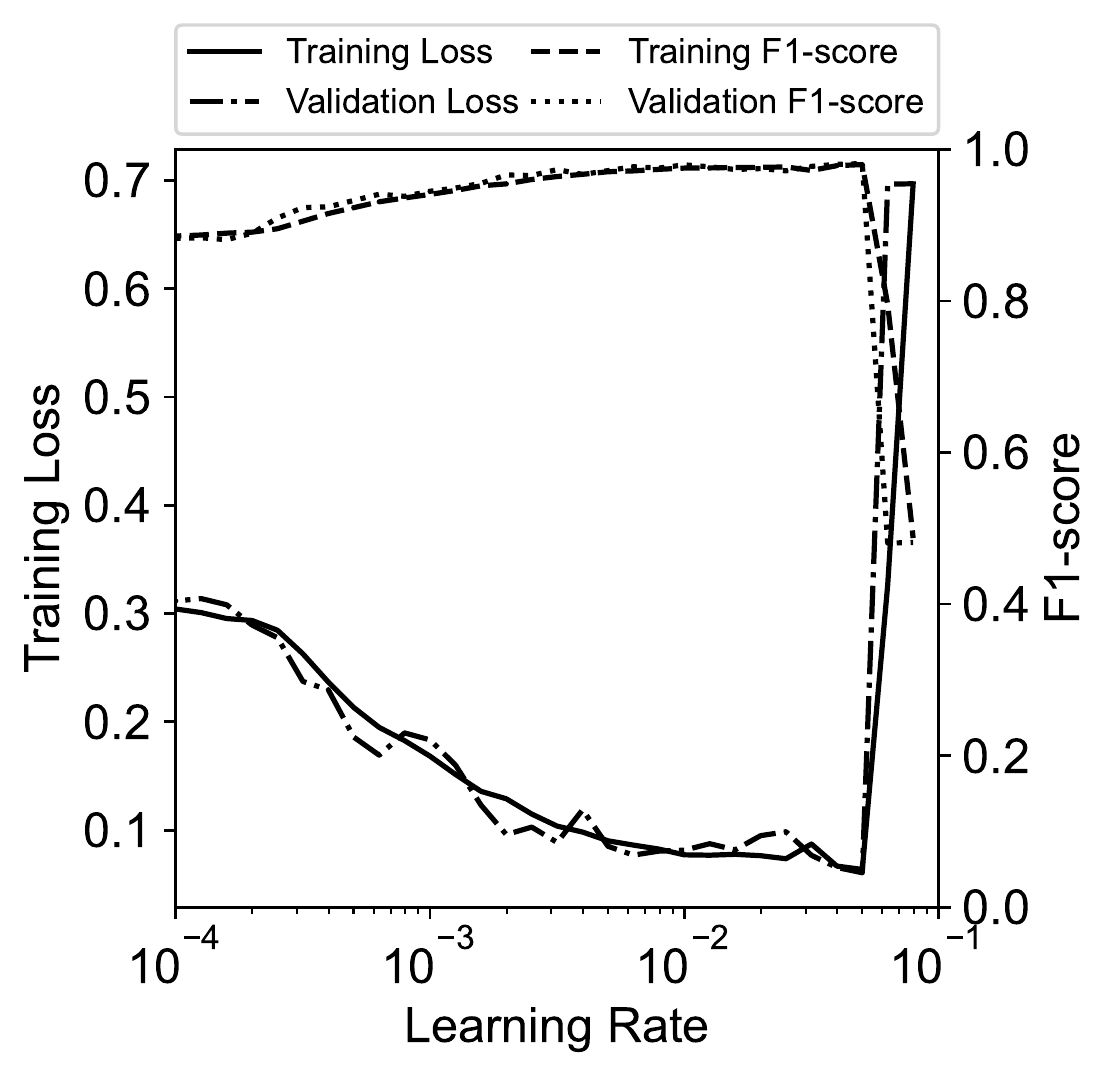}
    \subcaption{Loss}
    \label{fig:m6lrloss}
  \end{minipage}
  \begin{minipage}[b]{0.49\columnwidth}
  \centering
    \includegraphics[scale=0.40,keepaspectratio]{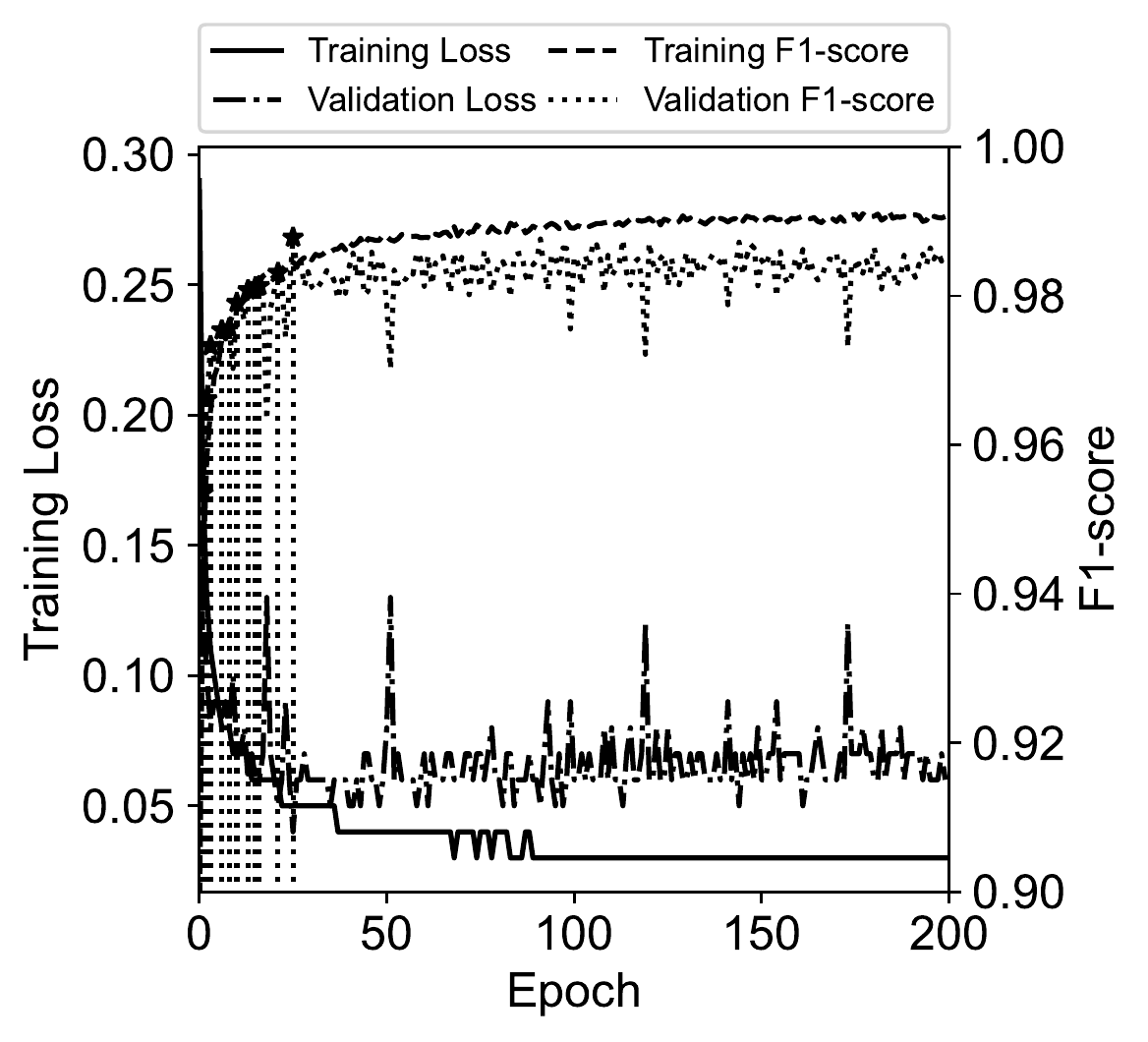}
    \subcaption{Accuracy}
    \label{fig:m6lraccu}
  \end{minipage}
 \caption{HM6 DNN performance (loss and f1-score) under different learning rates and the optimal value of $\sim$0.0063 with the accuracy of 98.56\%.
 }
 \label{fig:m6}
\end{figure}

The helper models also go through hyperparameters optimization for the desired learning rates. Figure \ref{fig:m1} presents \emph{HM1} depicting the corresponding optimal learning rate and the model performance. In this model, the optimal learning rate is equal to $10^{-4}$, achieving the model accuracy of 95.86\% for the test datasets. Adding new features for \emph{HM2} in Figure \ref{fig:m2} led to achieving an accuracy of 95.39\% for the obtained optimal learning rate equal to $10^{-3}$. The designed CNN model with the given feature for \emph{HM3}\footnote{The optimal learning rates for helper models HM3, HM5, and HM6, however, were not led to training. Thus, we experimentally evaluated the next top learning rates that would result in the training of the models with the next optimal ones.} shows an accuracy of 98.88\% for the learning rate of $0.00398$ in Figure \ref{fig:m3}, presenting an improvement of more than 3\% compared to the MLP-based models. Compared to \emph{HM2}, Figure \ref{fig:m4} presents \emph{HM4} with higher accuracy for the obtained learning rate of $5 \times 10^{-5}$ and corresponding accuracy of 99.23\%, which is higher than the other MLP-based models. For more complex feature selections (i.e., \emph{HM5} and \emph{HM6}, the CNNs achieved 98.99\% (Figure \ref{fig:m5}) and 98.56\% (Figure \ref{fig:m6}) accuracy with respect to corresponding optimal learning rates, $2.511 \times 10^{-3}$ and $6.3 \times 10^{-3}$.

Hence, incorporating more features will lead to higher accuracy based on the designed model, concluding that CNN-based models are relatively achieved the highest accuracy. 

All the feature-specific DNN models are then converted to mobile-friendly formats (i.e., Tensorflow lite (.tflite)) and transferred to smartphones. In addition, the ADAM collaborative model is prepared, colocating with the models on smartphones for collaborative learning and detection.

\subsection{Experiment Results}

Seven recent representative smartphones, updated to the latest official versions (Android 12 and 13), are employed for evaluation: Motorola G62 5G 128GB, Samsung Galaxy S20 FE 5G 128GB, Google Pixel 6a 5G 128GB, OPPO Find X3 Lite 5G 128GB, OPPO Reno8 Lite 5G 128GB, Xiaomi Pad 5 128GB, and OPPO Find X3 Neo 5G 256GB. Smartphones run benign, malware, or a mixture of both applications, simulating a diverse experiment setup. In addition, these phones employ a combination of more than 700 applications installed (published in the past three quarters of 2022) from Google Play and third-party repositories belonging to benign or malware classes. Smartphones may have random duplicate applications installed to evaluate the sensitivity of collaborative DNN models on smartphones when trained with on-device features.

Evaluation results are discussed as follows. ADAM initially presents the predicted labels for the installed applications across smartphones by employing feature-specific models to set the labels for the unlabeled application. In the following, ADAM initially employs smartphone vendor applications, referred to as operating system applications tagged as ``system'' applications, for training the collaborative DNN models. In addition to a test dataset of 1000 applications, such applications train federated learning guards to protect the collaborative model from poisoning. Moreover, \emph{pseudo labeling} technique is used to retrain the collaborative DNN models. Eventually, the collaborative malware detection for improving the collaborative DNN model is done through federated learning, aggregating the training parameters, and employing the federated learning guards to avoid poisoning the collaborative model.

Smartphone applications are analyzed for the manifest file and the Dex class(es) to extract the features for inference and training. Features per each application are stored as float buffer array files on the smartphone, maintaining the extracted features for sharing and further processing. In particular, for features regarding the Dex class(es), per each one, the following are used for the corresponding API classes and the sensitive features. According to the library \cite{dexlib2}, corresponding method-level features are extracted by employing ``getFieldSection()'' and ``getMethodSection()'', which provide not only the method-level ones but also the corresponding used APIs. In addition, APIs are extracted through ``getclasses()'', ``getProtoSection()'', ``getTypeRefernces()'' (or ``getStringReferences()'') which are equivalent to the ones used in Section \ref{sec:problem_statement}.

\begin{figure}[!t]
  \centering
    \begin{minipage}[b]{0.49\columnwidth}
    \centering
    \includegraphics[scale=0.41,keepaspectratio]{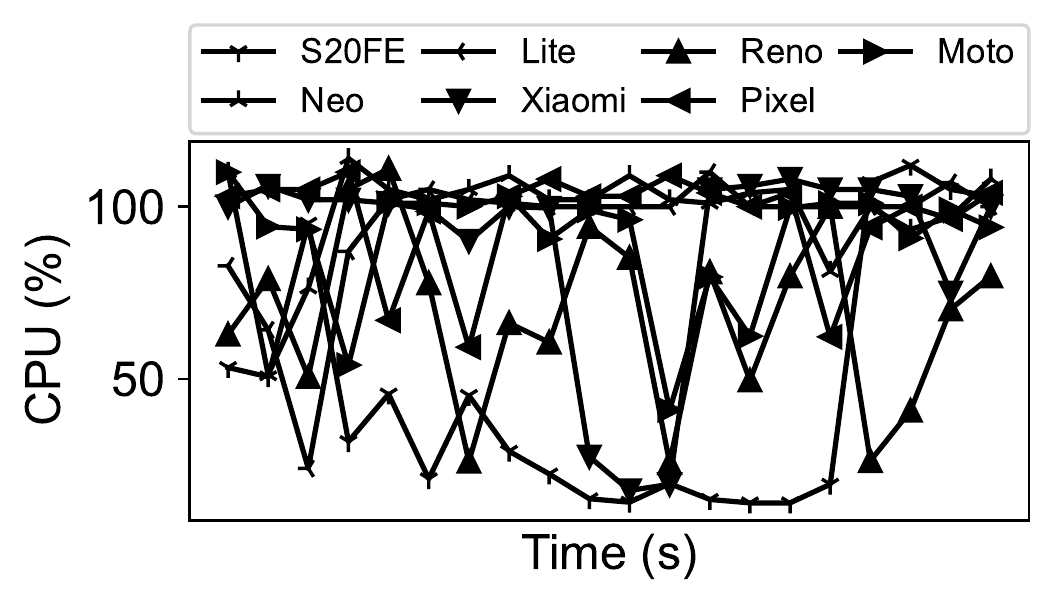}
    \subcaption{CPU utilization}
    \label{fig:mob_anatra}
  \end{minipage}  
  \begin{minipage}[b]{0.49\columnwidth}
  \centering
    \includegraphics[scale=0.41,keepaspectratio]{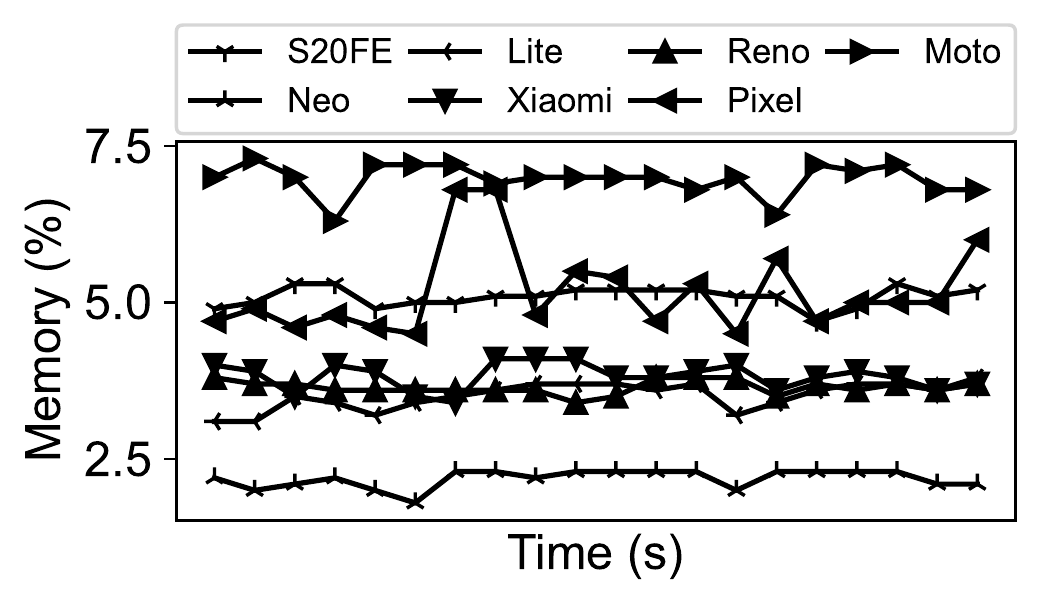}
    \subcaption{Memory utilization}
    \label{fig:mob_anatrb}
  \end{minipage}
 \caption{Resource utilization of application analysis for different smartphones. Valleys represent being busy with unpacking the application and saving it in storage.} 
 \label{fig:mob_anatr}
\end{figure}

ADAM application is gone through the resource utilization (\%) across different smartphones shown in Figure \ref{fig:mob_anatr} throughout the monitoring period for 250 seconds as samples are collected per 5 seconds interval. ADAM considers the system applications (i.e., the Android operating system and the vendor applications) for processing analysis. Due to diversity, such resource utilization is considered representative of the ADAM application.  

ADAM collects relevant features for the feature-specific DNN models. We exclude applications in which Dex classes and the manifest file are unavailable, resulting in the final analysis shown in Table \ref{tab:mobsys_info}.

\begin{table}[b!]
    \centering
    \caption{Smartphone System Applications}
    \begin{tabular}{|p{3cm}<{\centering}|p{1.5cm}<{\centering}|p{1.5cm}<{\centering}|p{1.1cm}<{\centering}|}
    \hline
        Smartphone & System applications & Processed & Non-processed \\
    \hline
    Motorola G62 & 335 & 237 & 98\\ \hline
    Samsung S20 FE & 410 & 363 & 47  \\ \hline
    Google Pixel 6a & 267 & 187 & 80\\ \hline
    OPPO Find X3 Lite & 329 & 271 & 60 \\ \hline
    OPPO Reno8 Lite & 359 & 257 & 102 \\ \hline
    Xiaomi Pad 5 & 312 & 225 & 87 \\ \hline
    OPPO Find X3 NEO & 329 & 269 &  58 \\ \hline

    \end{tabular}
    \label{tab:mobsys_info}
\end{table}

\subsection{Feature-Specific DNN Models Efficiency}
\label{sec:model_efficiency}

\subsubsection{Application Fingerprint Sensitivity} 
In Section \ref{sec:model_generic}, we argued that even with the entire application fingerprint, there is a probability that an application might be wrongly labeled. This section empirically shows how different feature-specific DNN models can assist with more confident labeling.

\begin{table}[h!]
    \centering
    \caption{Benign ($\mathbb{P}_\beta$)/Malware ($\mathbb{P}_\mu$) Confidence Probability}
    \begin{tabular}{|p{0.1cm}<{\centering}|p{0.76cm}<{\centering}|p{0.76cm}<{\centering}|p{0.76cm}<{\centering}|p{0.76cm}<{\centering}|p{0.76cm}<{\centering}|p{0.76cm}<{\centering}|p{0.76cm}<{\centering}|}
    \hline
        \# & Static & HM1 & HM2 & HM3 & HM4 & HM5 & HM6 \\
    \hline
    1 &$\mathbb{P}_\beta$=0 \; $\mathbb{P}_\mu$=1 &  $\mathbb{P}_\beta$=1 \; $\mathbb{P}_\mu$=0 & 
        $\mathbb{P}_\beta$=0.99  $\mathbb{P}_\mu$=0.01 &
        $\mathbb{P}_\beta$=1 \; $\mathbb{P}_\mu$=0 &
    $\mathbb{P}_\beta$=0.99  $\mathbb{P}_\mu$=0.01 & 
    $\mathbb{P}_\beta$=1 \; $\mathbb{P}_\mu$=0&
    $\mathbb{P}_\beta$=0.99  $\mathbb{P}_\mu$=0.01 \\ \hline
    2 & $\mathbb{P}_\beta$=0 \; $\mathbb{P}_\mu$=1 & $\mathbb{P}_\beta$=0.9  $\mathbb{P}_\mu$=0.1 & $\mathbb{P}_\beta$=0.99  $\mathbb{P}_\mu$=0.01 &
    $\mathbb{P}_\beta$=0 \; $\mathbb{P}_\mu$=1 & $\mathbb{P}_\beta$=0 \;$\mathbb{P}_\mu$=1& $\mathbb{P}_\beta$=0 \;$\mathbb{P}_\mu$=1&
    $\mathbb{P}_\beta$=0.02 $\mathbb{P}_\mu$=0.98 \\ \hline

    \end{tabular}
    \label{tab:twoappana}
\end{table}

\begin{figure}[!b]
  \centering
    \begin{minipage}[b]{0.49\columnwidth}
    \centering
    \includegraphics[scale=0.4,keepaspectratio]{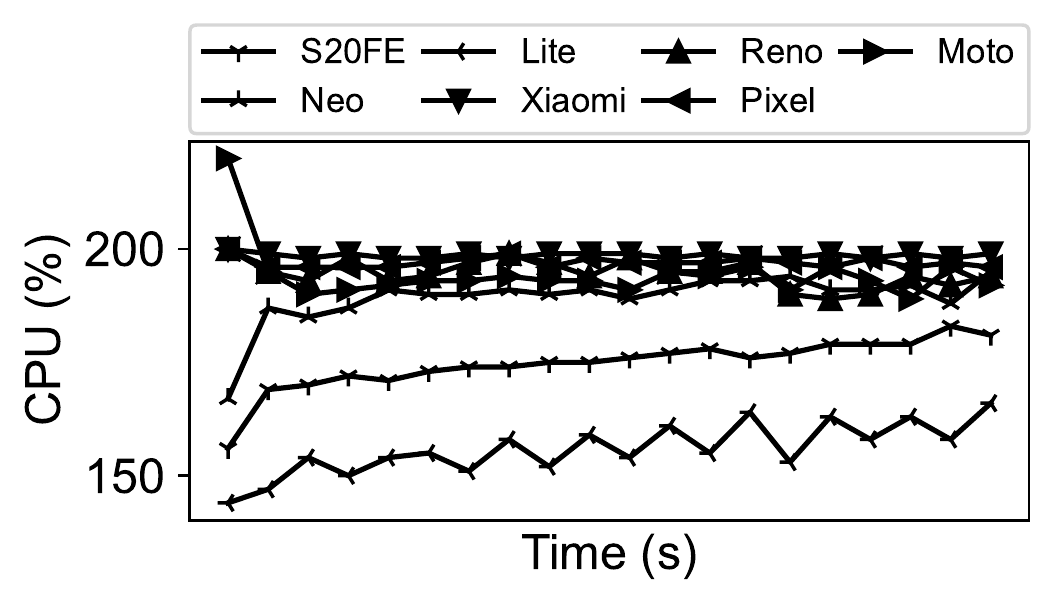}
    \subcaption{}
    \label{fig:mob_anatra}
  \end{minipage}  
  \begin{minipage}[b]{0.49\columnwidth}
  \centering
    \includegraphics[scale=0.4,keepaspectratio]{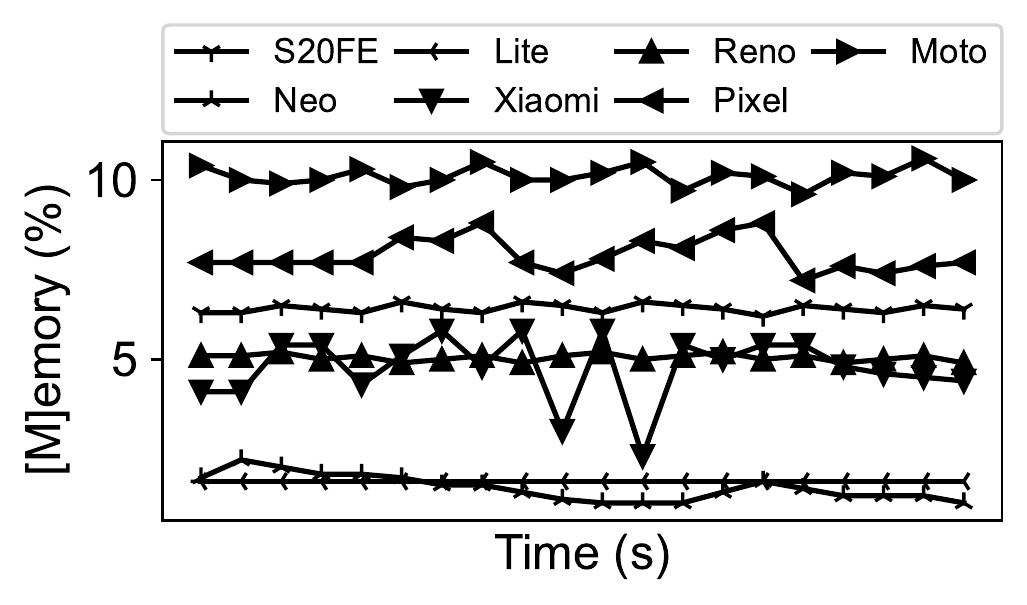}
    \subcaption{}
    \label{fig:mob_anatrb}
  \end{minipage}
  
    \begin{minipage}[b]{1\columnwidth}
    \centering
    \includegraphics[scale=0.415,keepaspectratio]{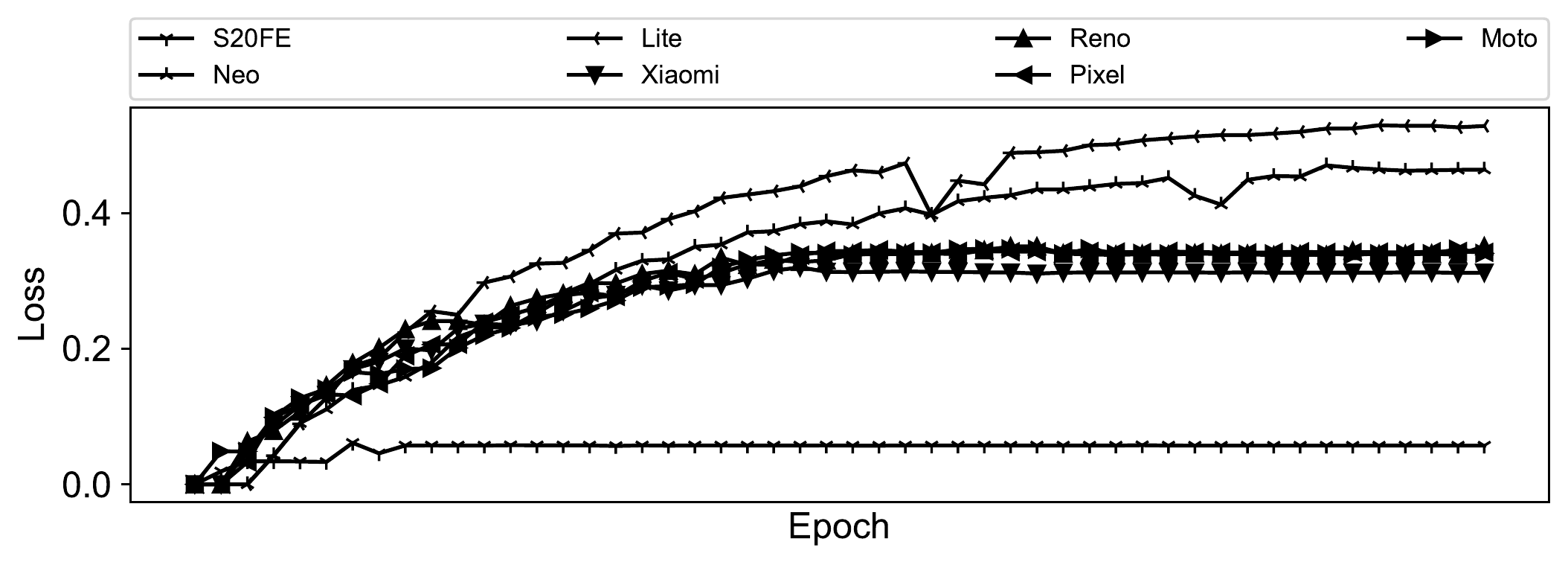}
    \subcaption{}
    \label{fig:mob_anatrc}
  \end{minipage}

 \caption{ADAM training resource utilization and loss for ADAM's collaborative model.} 
 \label{fig:mob_anatrp}
\end{figure}

We install the ``Google Authenticator'' application (\#1) through Google Play on a smartphone and use the full application fingerprint and the associated subset of features (i.e., helper models) to label the application. This application is chosen because users widely use it to facilitate multi-factor authentication, and many companies trust it to protect users' accounts from unauthorized access. Hence, it is a ``benign'' application. Also, a shop interface application (\#2) already labeled as malware by VirusTotal is analyzed, and the ``confidence'' probability is provided, rounded to two decimal digits for simplicity. In Table \ref{tab:twoappana}, Google Authenticator is wrongly labeled as malware by employing the full application fingerprint, while the helper models assisted with the correct label. Conversely, the shop interface is correctly labeled as malware by the full fingerprint but wrongly labeled by the associated helper models, HM1 and HM2. This implies selecting correct \emph{features} for labeling is essential, incorporating API classes and/or their corresponding methods and without considering the pre-defined features not filled in the manifest. This becomes important when empty manifest files (i.e., only parent tag elements) are investigated, such as the font typeface Rosemary \cite{rosemarytypeface} installed on Samsung S20FE OS, resulting in the ``malware'' label for all the feature-specific DNN models. Similar to Figure \ref{fig:mob_anatr}, in addition to an overall loss for training, ADAM resource utilization has increased due to training and handling more features shown in Figure \ref{fig:mob_anatrp}.  

\begin{figure}[!h]
  \centering
    \begin{minipage}[b]{0.49\columnwidth}
    \centering
    \includegraphics[scale=0.39,keepaspectratio]{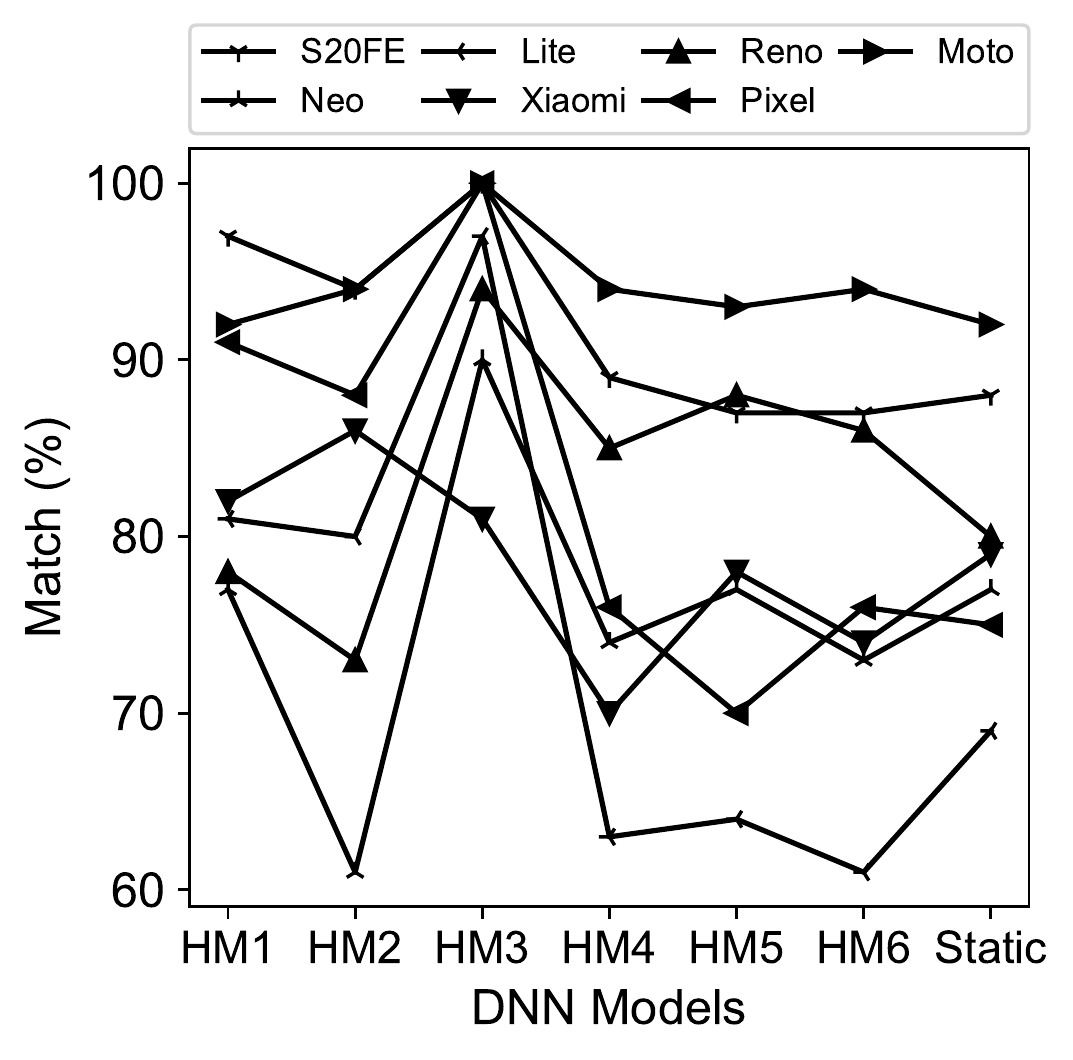}
    \subcaption{Pseudo labeling match}
    \label{fig:adamconfg}
  \end{minipage}
  \begin{minipage}[b]{0.49\columnwidth}
  \centering
    \includegraphics[scale=0.39,keepaspectratio]{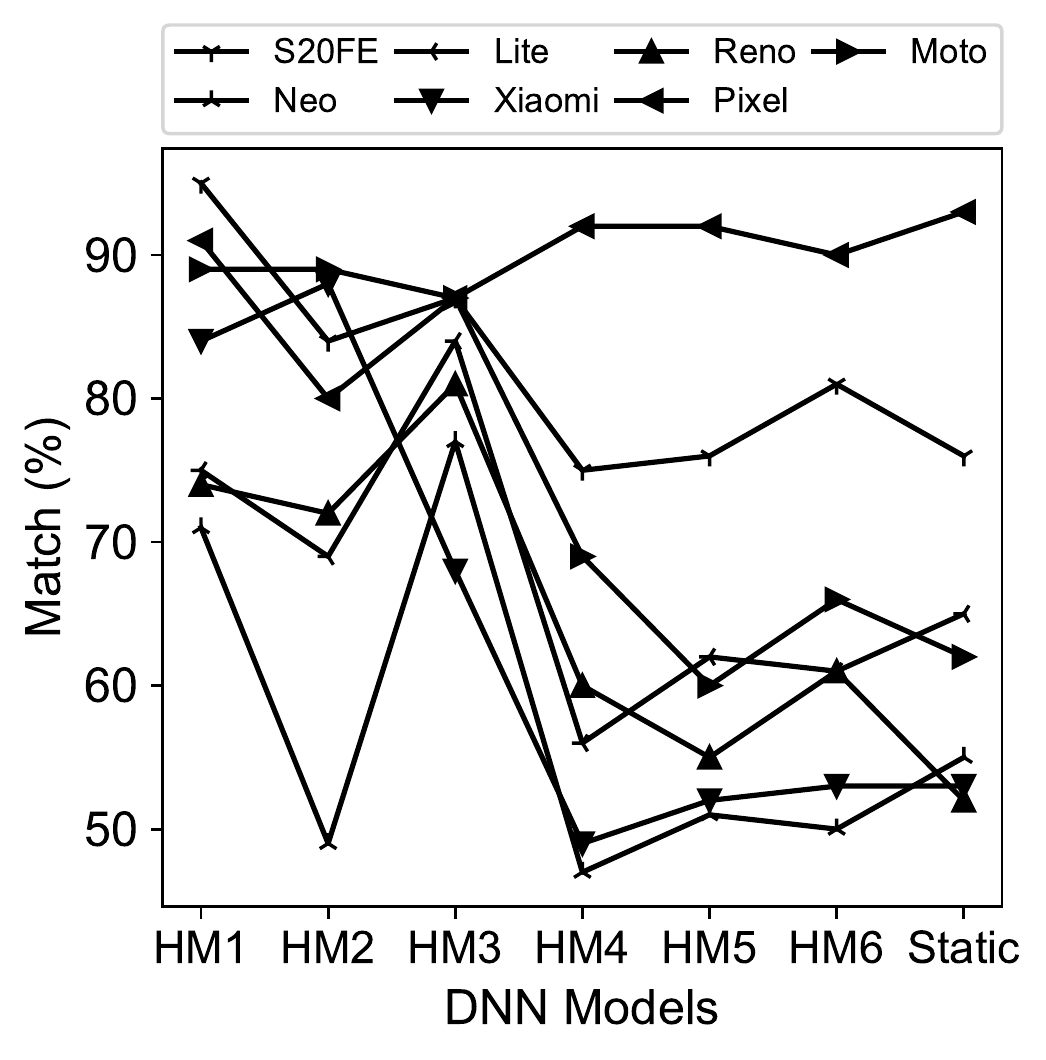}
    \subcaption{Ground truth label match}
    \label{fig:adamconfy}
  \end{minipage}

 \caption{Pseudo labeling and its comparison to the ground truth efficiency by feature-specific models.}
 \label{fig:hm_eff}
\end{figure}

Figure \ref{fig:hm_eff} shows the final labels produced by all feature-specific labels for the pseudo labeling (\ref{fig:adamconfg}) and its comparison with the ground truth (\ref{fig:adamconfy}). Specifically, it shows not only the feature-specific DNN models \emph{inference} accuracy but also the comparison with the \emph{actual} labels. 
By comparing the models, HM3 has achieved the top accuracy concerning the pseudo and actual labels. The other top models are reported as HM1, HM6, HM4, HM5, Static, and HM2, across the smartphones, resulting in the upper bound accuracy level of more than 70\% for the collaborative model compared to the poor performance of smartphones' (or vendor) antivirus and Google Play Protect reported in \cite{pasdardom}.

\subsubsection{ADAM Collaborative Models Preparation}
Extracted features and the corresponding labels provided by the feature-specific DNN models are used to build training datasets for collaborative DNN models. In addition, such datasets are used to train an \emph{initial} DNN for smartphones at the federated learning server and assist with equipping federated learning guards, which will be discussed in the following section. These collaborative learning models are CNN-based, the static model, and three helper models, HM3, HM5, and HM6. 

The initial collaborative DNN models are based on the OS/vendor applications where they are tagged as ``system'' in the Android OS. In addition, a small set of benign applications (i.e., 1000 unseen apps) are mixed with those applications to prepare each collaborative DNN model. Altogether these applications build a validation set to assess models at the federated learning server. Except for MLP-based models, CNN-based models are enabled with the ``base'' and ``head'' parts, allowing them to freeze the first part, attach similar ANN networks, and train them with these validation samples obtained from smartphones. Weights of the ANNs are then extracted and sent to smartphones for employing as weight \emph{initialization} of collaborative DNN models.

\begin{figure}[!t]
  \centering
    \begin{minipage}[b]{0.49\columnwidth}
    \centering
    \includegraphics[scale=0.385,keepaspectratio]{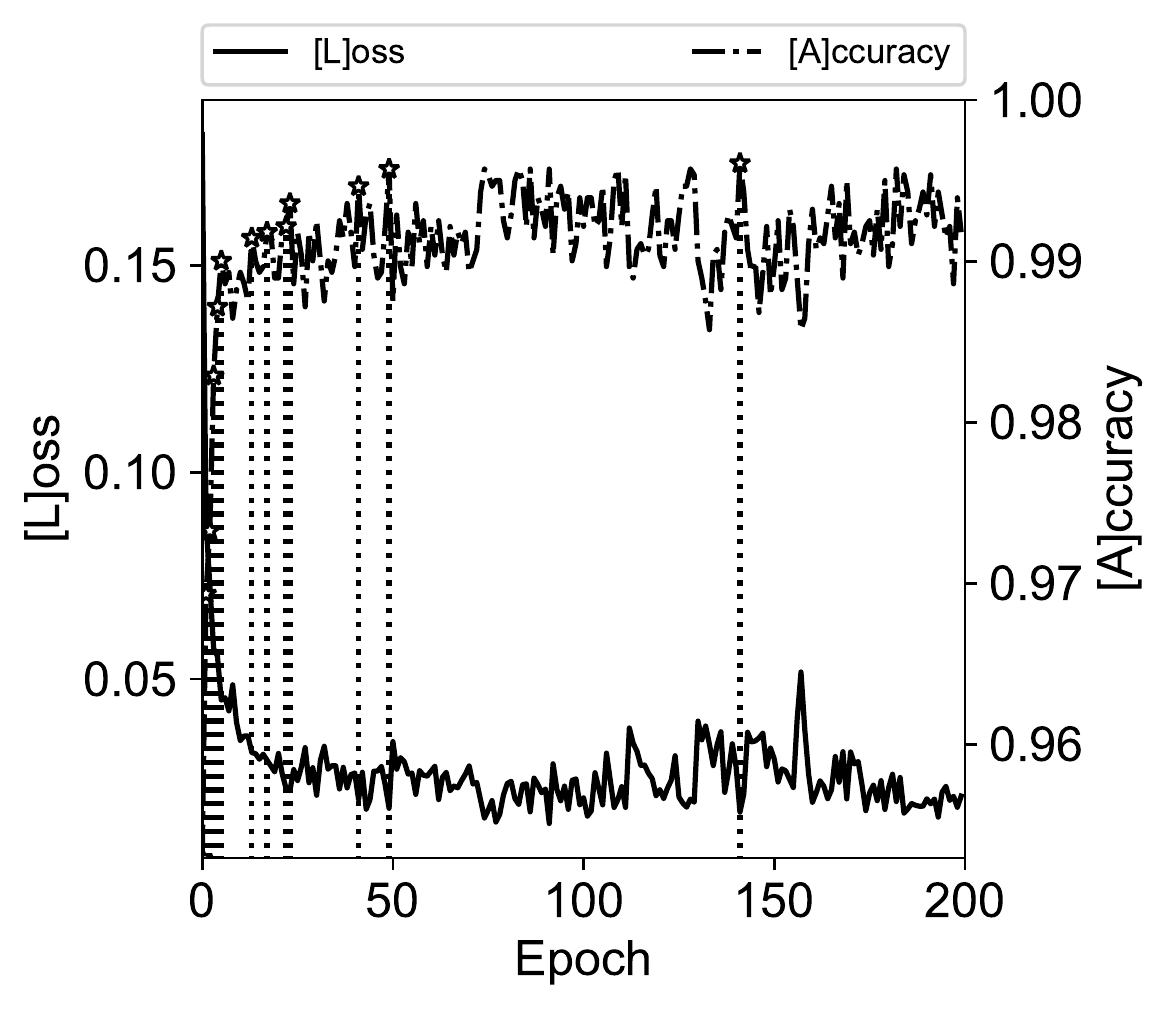}
    \subcaption{Static}
    \label{fig:mob_flguardsa}
  \end{minipage}  
  \begin{minipage}[b]{0.49\columnwidth}
  \centering
    \includegraphics[scale=0.385,keepaspectratio]{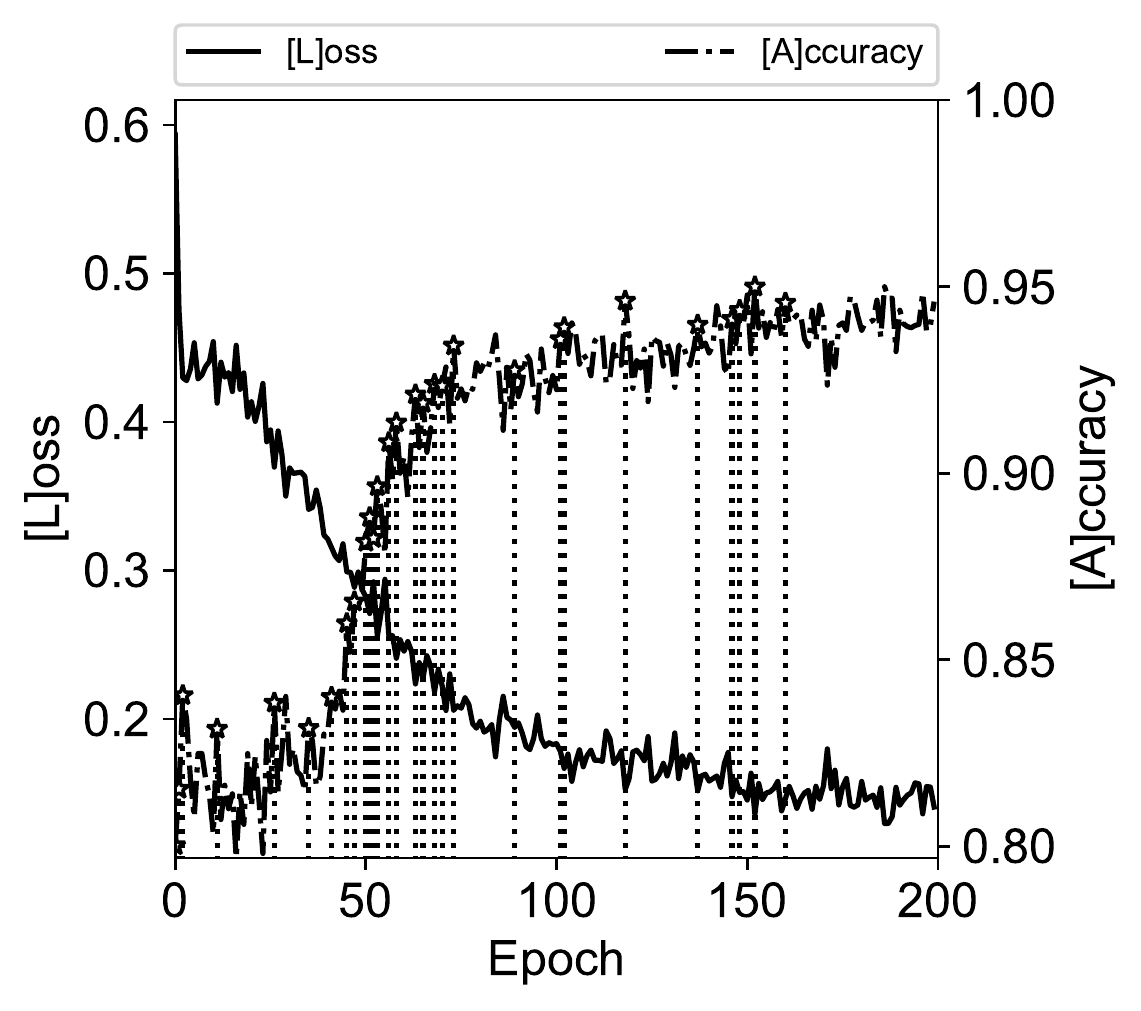}
    \subcaption{HM3}
    \label{fig:mob_flguardsb}
  \end{minipage}
    \centering
    \begin{minipage}[b]{0.49\columnwidth}
    \centering
    \includegraphics[scale=0.39,keepaspectratio]{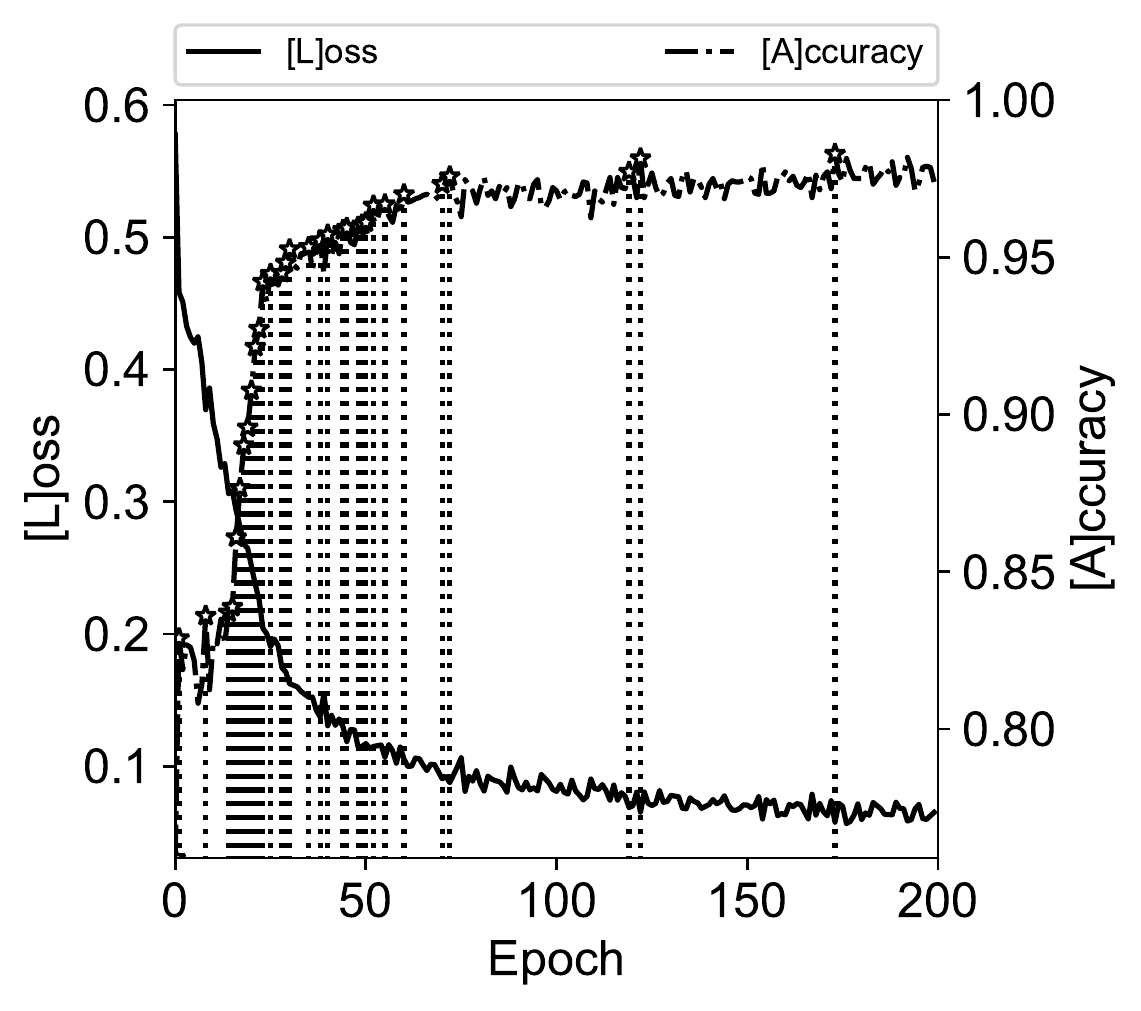}
    \subcaption{HM5}
    \label{fig:flguardsc}
  \end{minipage}  
  \begin{minipage}[b]{0.49\columnwidth}
  \centering
    \includegraphics[scale=0.39,keepaspectratio]{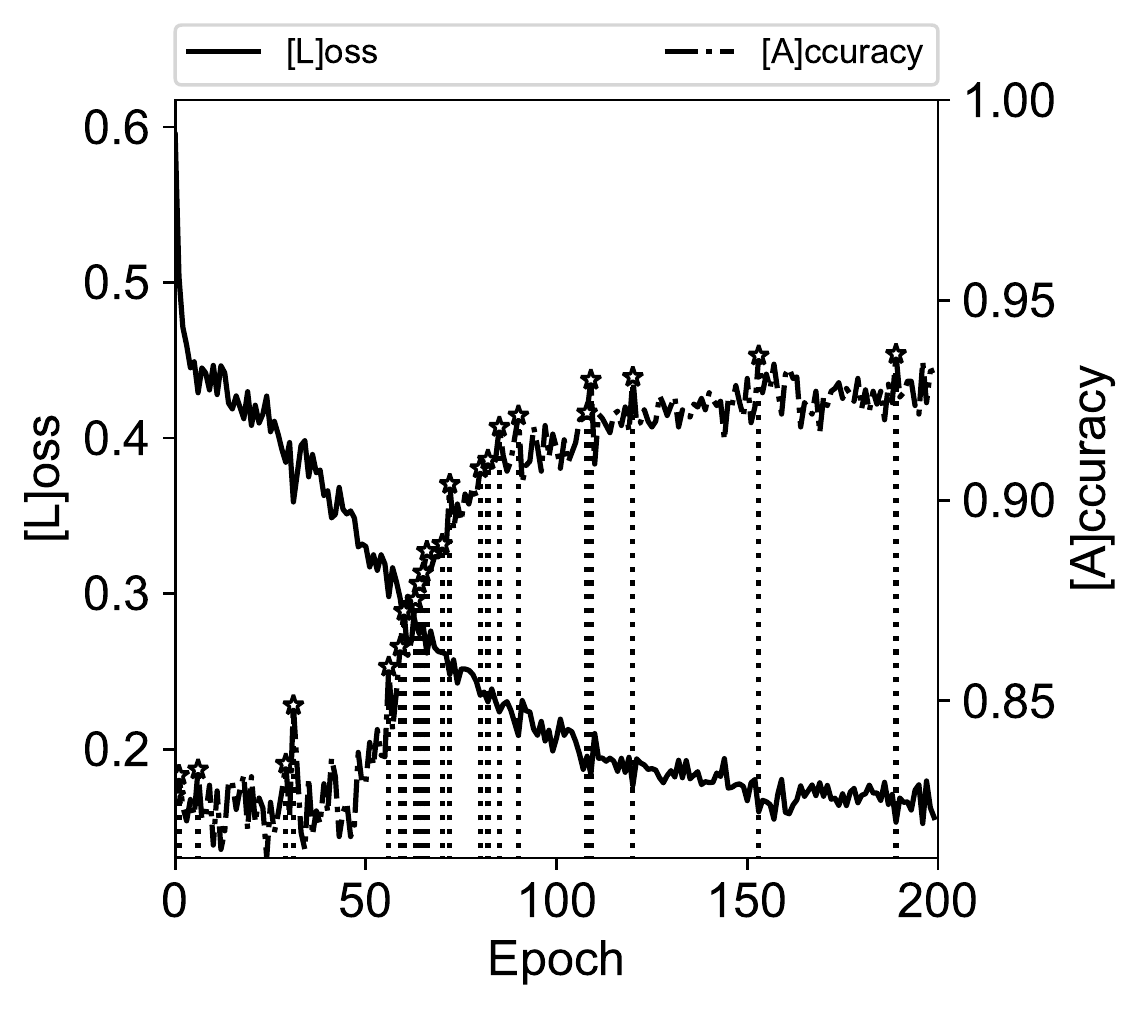}
    \subcaption{HM6}
    \label{fig:flguardsd}
  \end{minipage}

 \caption{Federated learning guards performance (loss and accuracy). Dotted vertical lines represent a model improvement.} 
 \label{fig:flguards}
\end{figure}

These validation samples prevent the collaborative models from being poisoned across smartphones. Hence, we define federated learning model guards to actively participate in monitoring the models before aggregating them on the server. The performance of these guard models is shown in Figure \ref{fig:flguards}. In terms of accuracy and loss, the federated learning model guards for the static-based one achieved higher accuracy (i.e., more than $\sim$99.20) compared to $\sim$94\%, $\sim$97\%, and $\sim$94\% for HM3, HM5, and HM6, respectively.

\subsection{ADAM Federated Learning Model Guards Efficiency}

In this section, we evaluate the robustness of the collaborative DNN models against poisoning attacks via the help of federated learning model guards. The weight manipulation attacks are about altering the collaborative DNN model weights (i.e., classification layers) to reduce the model's performance across smartphones. The labels of smartphone training datasets, in label-flipping attacks, are manipulated to inject misleading labels into the models and decrease the performance of the collaborative DNN models.

\begin{figure}[!h]
  \centering

  \begin{minipage}[b]{1\columnwidth}
    \centering
    \includegraphics[scale=0.41,keepaspectratio]{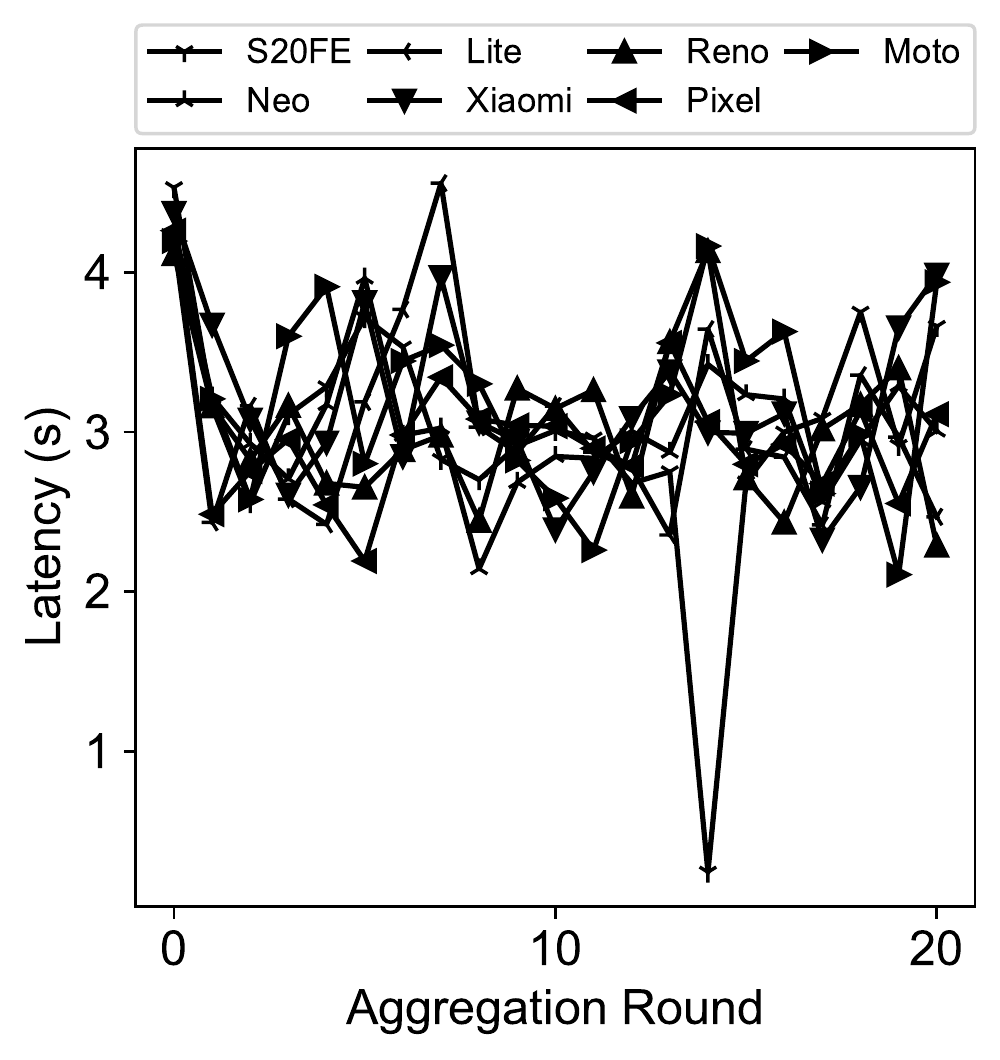}
    \subcaption{Communication latency}
    \label{fig:flguardeffia}
  \end{minipage}
    \begin{minipage}[b]{0.49\columnwidth}
    \centering
    \includegraphics[scale=0.41,keepaspectratio]{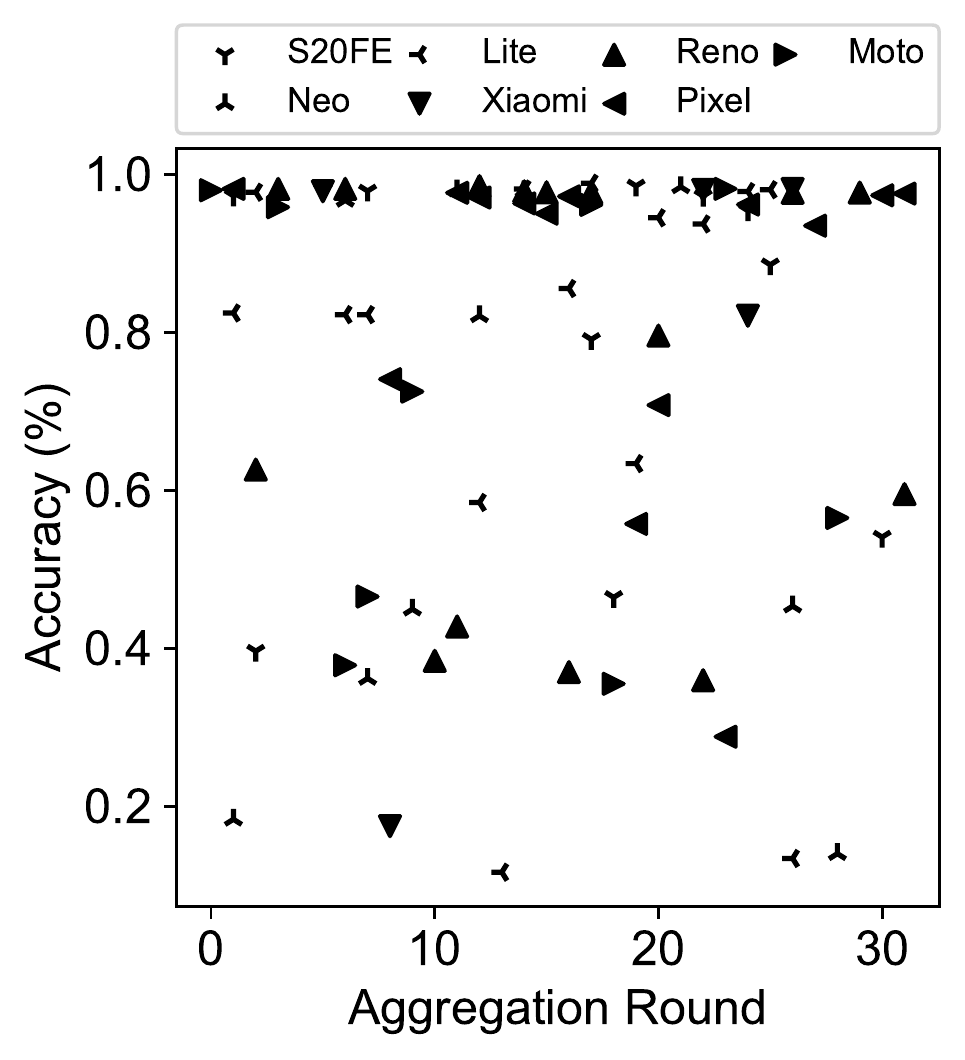}
    \subcaption{Accuracy}
    \label{fig:flguardeffib}
  \end{minipage}
  \begin{minipage}[b]{0.49\columnwidth}
  \centering
    \includegraphics[scale=0.41,keepaspectratio]{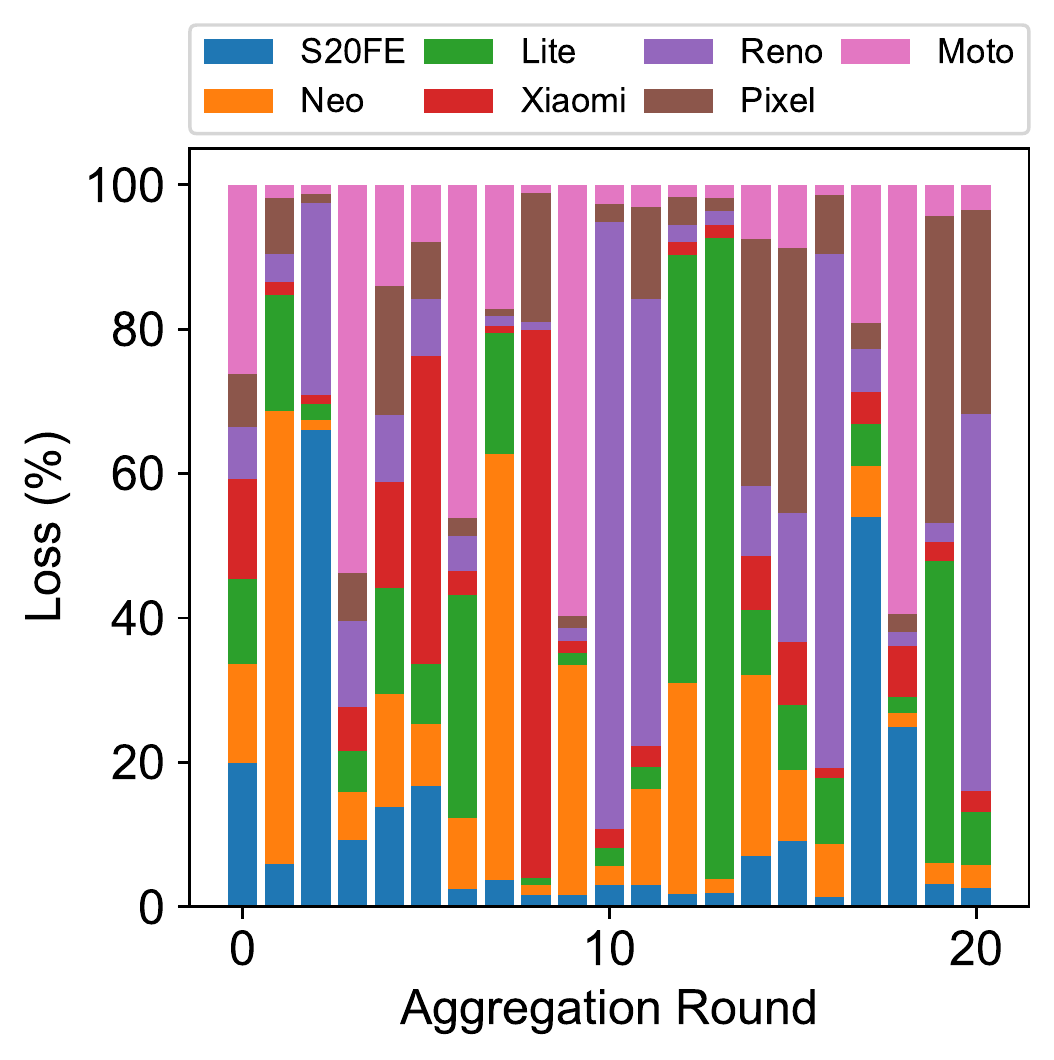}
    \subcaption{Loss percentage}
    \label{fig:flguardeffic}
  \end{minipage}
  \begin{minipage}[b]{0.49\columnwidth}
    \centering
    \includegraphics[scale=0.41,keepaspectratio]{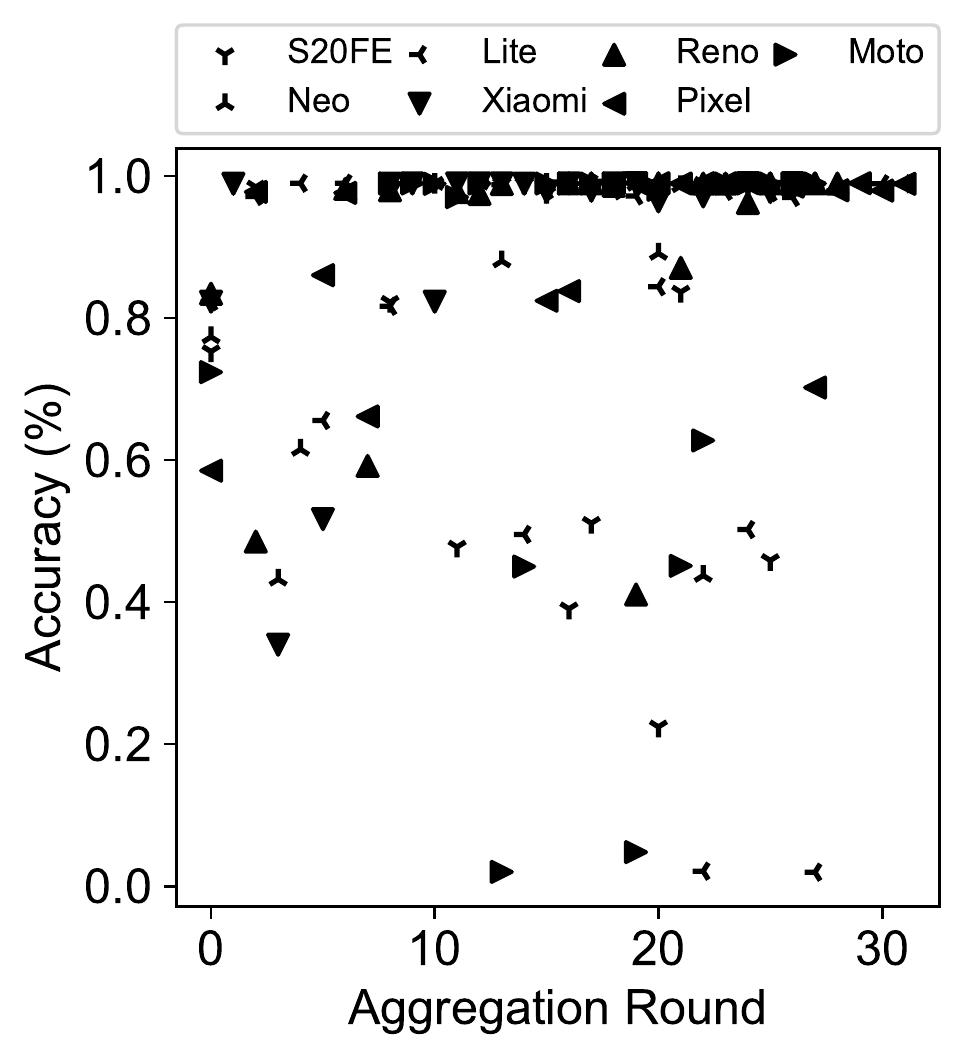}
    \subcaption{Accuracy}
    \label{fig:flguardeffid}
  \end{minipage}
  \begin{minipage}[b]{0.49\columnwidth}
    \centering
    \includegraphics[scale=0.41,keepaspectratio]{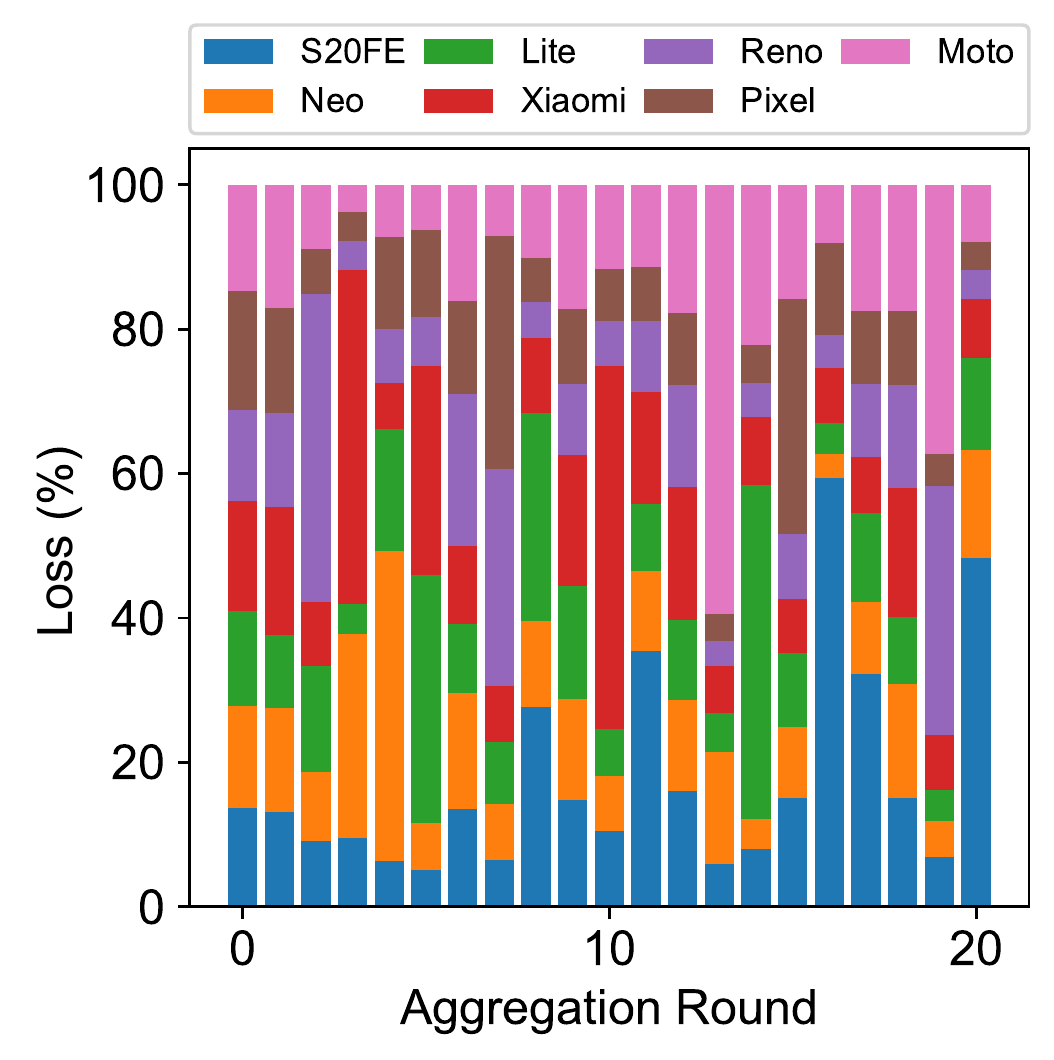}
    \subcaption{Loss percentage}
    \label{fig:flguardeffie}
  \end{minipage}

 \caption{Communication latency (a) and federated learning guards efficiency to exclude poisonous weights at different aggregation rounds for (b,c) OS/vendor applications and (d,e) with user applications. Non-equal blocks represent the proportion of (bigger/smaller) loss in each round.}
 \label{fig:flguardeffi}
\end{figure}

\subsubsection{Model Weight Manipulation}

Collaborative DNN models are evaluated against weight manipulation by altering their parameters (i.e., weights) across smartphones. Per each aggregation rounds, federated learning model guards evaluate weights before aggregation to filter \emph{non-compliance} and \emph{malicious} participants from involving in the weight updates. Smartphones are \emph{randomly} selected to diversify the choice of participants, and weights of each malicious participant (i.e., smartphone) are manipulated with Equation \ref{eq:weight_man}, which involves partial or complete weight manipulation based on the two parameters; $lb$ and $ub$. The former is the lower bound, while the latter is the upper bound of weights indexes, and $U$ represents the uniform distribution of numbers between $-1$ and $1$.

\begin{equation}
    \label{eq:weight_man}
    W_{\omega}[lb:ub] = W_{\omega}[lb:ub] * (U[-1,1] * |ub-lb|).
\end{equation}

For weight manipulation, the federated learning guards check each smartphone's parameters by \emph{loading} their training parameters at the classification layers, and the DNN performance is checked in terms of loss, recall/accuracy. Each federated learning guard is assigned a certain accuracy threshold to filter non-compliance smartphones if the results are under the threshold to exclude the smartphone from the aggregation round; otherwise, the aggregated model is returned. 

We first present the communication latency between the federated learning server and smartphones in addition to the accuracy and loss evaluation shown in Figure \ref{fig:flguardeffi}. Also, reloading the federated learning guards (GPU-based) with the smartphone weights and its assessment against the evaluation datasets takes $\sim$30s. This latency is bounded due to involving either CPU or GPU. Then, the sensitivity of the federated learning guards is evaluated with only OS/vendor applications and including user applications.

Figure \ref{fig:flguardeffia} shows that for a model with a size of $338$ KiB, the average latency, including the underlying OS delays and transmission latency, is near 3.5s, choosing 5s as the upper bound delay at the federated learning server for each aggregation. Figure \ref{fig:flguardeffib} shows the excluded smartphones due to violating the threshold for consecutive aggregation rounds, emphasizing that federated learning model guards are effective and resilient to weight manipulation. In addition, if the average weight resulted in $NaN$ (i.e., not a number) in the aggregation round, the federated learning server lets participants utilize the last \emph{usable} weights for continuous training. Figure \ref{fig:flguardeffic} shows the \emph{percentage} of loss values as stacked bar charts for only 20 aggregation rounds for visibility. This figure illustrates that the loss value of smartphones is affected by weight manipulation, in which corresponding accuracy is used to filter the poisonous ones. Figures \ref{fig:flguardeffid} and \ref{fig:flguardeffie} present the weight manipulation effect when the pseudo-labeling technique is applied in the presence of user applications in addition to the system apps. Compared to Figure \ref{fig:flguardeffib}, the threshold mechanism filtered more non-compliance smartphone weights due to updated collaborative model parameters with the smartphone user applications. Also, the stacked loss par charts compared to Figure \ref{fig:flguardeffic} presents more consistency in loss changes.  

\subsubsection{Label Flipping and Application Features Manipulation}

Collaborative DNN models are also assessed against label-flipping attacks by altering the class of an application it belongs to, i.e., malware to benign and vice versa, or application features manipulation. In the former, this alteration requires injecting the wrong label and its corresponding features into the training to reduce the collaborative models' performance eventually. In the latter, the label-flipping attacks \emph{may} require manipulation of the application features (i.e., changing a partial part of the fingerprint to $0$ or $1$) to drive the feature-specific DNN models to produce the desired label as ADAM sets the label based on the consensus among the models, i.e., more than $N/2$  must report the desired label. 

In each round of aggregation and in addition to the collaborative DNN model weights, federated learning guards request the full \emph{user} applications fingerprint. As the guards are already trained with the OS/vendor applications, this results in mitigating the impact on the system-based applications, shown in Figures \ref{fig:flguardeffib} and \ref{fig:flguardeffic}. In addition, the consensus among the guards will detect only flipping the application class, leading to mitigating the attacks. Federated learning guards are chosen based on Figure \ref{fig:adamconfy}, which have more consistency but diversity, i.e., Static, HM3, HM5, and HM6 as the monitoring models. In each round and prior aggregation\footnote{The full application fingerprint has a size of $161.7$ KiB, which can be down to $\sim$$1.3$ KiB, empirically resulting in negligible communication overhead.}, the guards vote on the \emph{inferred} labels and eventually compare them with the smartphone collaborative model parameters inference. This is done by reloading the guards with collaborative smartphone parameters to feed the fingerprint. If labels are the same, the smartphone's collaborative models are considered for aggregation; otherwise, they are omitted from the aggregation rounds. Equation \ref{eq:feature_man} presents how application features are manipulated.

\begin{equation}
    \label{eq:feature_man}
    \forall \; i \in \mathbb{F}[lb:ub] = \left\{
    \begin{array}{ll}
        1 & \; U[0,1] < 0.5 .\\
        0 & otherwise.
    \end{array}
\right. 
\end{equation}

\begin{figure}[!t]
  \centering
    \includegraphics[scale=0.415,keepaspectratio]{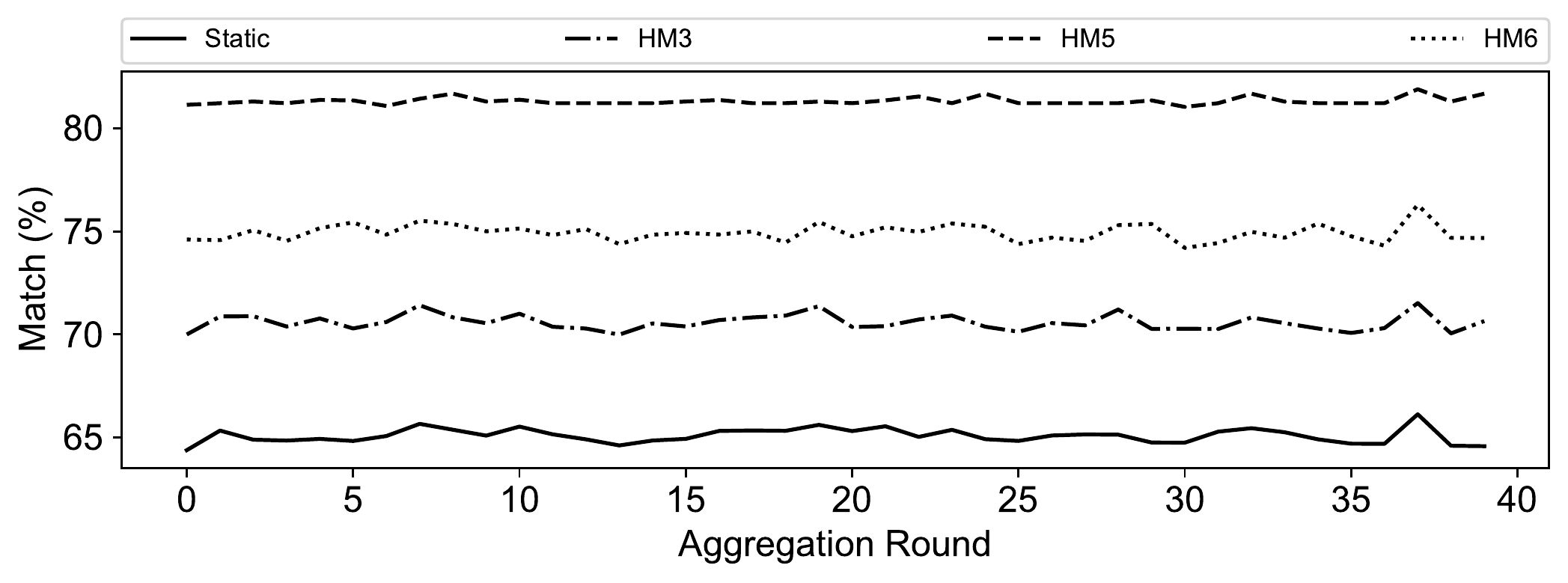}
    \label{fig:mob_anatrc}

 \caption{Feature manipulation and label flipping match accuracy.} 
 \label{fig:mob_fliplabel}
\end{figure}

Figure \ref{fig:mob_fliplabel} presents the average match-label accuracy for the selected four models across all seven smartphones. Each smartphone independently manipulates the application features (and labels) and sends them to the federated learning server for aggregation. This figure illustrates the sensitivity of feature manipulation and its final label inference in the federated learning side. Each DNN model shows different match accuracy, leading to realizing manipulation with the feature or labels. Compared to Figure \ref{fig:hm_eff}, HM3 achieved pretty similar matching accuracy to static, showing that the top model accuracy would eventually have the same performance. In contrast, the other two helper models had a much higher matching accuracy than their pseudo-labeling accuracy. 

\section{Conclusion}
\label{sec:conclusion}

We presented ADAM as an analytic-based malware detection that employs model personalization and transfer learning techniques. ADAM is built upon a main and a set of helper models, assisting with choosing high-confidence labels for applications with unknown ground truth. Each model is built and trained with respect to the most influential features, i.e., feature-specific models mentioned in the literature for malware detection, and is tailored to represent an application fingerprint. A comprehensive feature set dynamically in a privacy-preserving way is extracted on the device to label and train collaborative and universal DNN models across connected smartphones. Such models benefit from using federated learning for generalizing the models, being protected by federated learning guards to monitor and assess the quality of smartphones' collaborative parameters before each round of aggregation against data poisoning attacks. Evaluation results based on real-world applications show feature-specific models, on average, achieve 98\% accuracy, and the collaborative model has demonstrated outstanding performance against weight and feature (including the label) manipulation.

\bibliographystyle{IEEEtran}
\bibliography{main}

\end{document}